\documentclass[%
reprint,
superscriptaddress,
 amsmath,amssymb,
 aps,
prb,
longbibliography,
]{revtex4-2}

\usepackage{graphicx}
\usepackage{dcolumn}
\usepackage{bm}
\usepackage{enumitem}
\usepackage{siunitx}
\DeclareSIUnit\angstrom{\text{Å}}
\DeclareSIUnit\bar{bar}

\usepackage[final]{pdfpages} 
\usepackage{pgffor}

\usepackage[hidelinks]{hyperref}
\usepackage{xr}
\usepackage[capitalise]{cleveref}

\makeatletter
\AtBeginDocument{\let\LS@rot\@undefined}
\makeatother

\begin{document}

\newcommand*{\titlename}{Charge density waves and soft phonon evolution in the superconductor BaNi\textsubscript{2}(As\textsubscript{1-\textit{x}}P\textsubscript{\textit{x}})\textsubscript{2}}

\title{\titlename}
\author{Tom Lacmann}
\affiliation{Institute for Quantum Materials and Technologies, Karlsruhe Institute of Technology, 76021 Karlsruhe, Germany}
\author{Sofia-Michaela Souliou}
\affiliation{Institute for Quantum Materials and Technologies, Karlsruhe Institute of Technology, 76021 Karlsruhe, Germany}
\author{Fabian Henssler}
\affiliation{Institute for Quantum Materials and Technologies, Karlsruhe Institute of Technology, 76021 Karlsruhe, Germany}
\author{Mehdi Frachet}
\affiliation{Institute for Quantum Materials and Technologies, Karlsruhe Institute of Technology, 76021 Karlsruhe, Germany}
\author{Philippa Helen McGuinness}
\affiliation{Institute for Quantum Materials and Technologies, Karlsruhe Institute of Technology, 76021 Karlsruhe, Germany}
\author{Michael Merz}
\affiliation{Institute for Quantum Materials and Technologies, Karlsruhe Institute of Technology, 76021 Karlsruhe, Germany}
\affiliation{Karlsruhe Nano Micro Facility (KNMFi), Karlsruhe Institute of Technology, Kaiserstr. 12, 76131 Karlsruhe, Germany}
\author{Björn Wehinger}
\affiliation{ESRF, The European Synchrotron, 71, avenue des Martyrs, CS 40220 F-38043 Grenoble Cedex 9}
\author{Daniel A. Chaney}
\affiliation{ESRF, The European Synchrotron, 71, avenue des Martyrs, CS 40220 F-38043 Grenoble Cedex 9}
\author{Amir-Abbas Haghighirad}
\affiliation{Institute for Quantum Materials and Technologies, Karlsruhe Institute of Technology, 76021 Karlsruhe, Germany}
\author{Rolf Heid}
\affiliation{Institute for Quantum Materials and Technologies, Karlsruhe Institute of Technology, 76021 Karlsruhe, Germany}
\author{Matthieu Le Tacon}
\email{Matthieu.LeTacon@kit.edu}
\affiliation{Institute for Quantum Materials and Technologies, Karlsruhe Institute of Technology, 76021 Karlsruhe, Germany}

\date{\today}

\begin{abstract}
The superconductor BaNi\textsubscript{2}As\textsubscript{2} exhibits a soft-phonon-driven, incommensurate charge density wave (I-CDW) which is accompanied by a small orthorhombic structural phase transition. Upon further cooling, BaNi\textsubscript{2}As\textsubscript{2} undergoes a first-order structural transition to a triclinic phase in which a commensurate CDW (C-CDW) appears. The relationship and interplay between the I-CDW, C-CDW and structural phase transitions has remained elusive. To investigate this issue, we present a complementary study of thermal diffuse X-Ray scattering and inelastic X-Ray scattering for phosphorus substituted BaNi$_2$(As$_{1-\mathrm{x}}$P$_\mathrm{x}$)$_2$ \((x\lessapprox0.12\)) and down to \SI{2.2}{\kelvin}. We show that most of the diffuse scattering signal can be well described by first-principles lattice dynamics calculations. Furthermore, we find that although phosphorus substitution rapidly suppresses the structural transition temperatures, the temperature dependence of the correlation length of the I-CDW fluctuations and the formation of Bragg-like superstructure peaks associated with long-range ordering of this order depends only weakly on the substitution level.
Finally, we present the absence of signatures of the I-CDW to C-CDW or triclinic transition in the lattice dynamics, indicating that these instabilities are not (soft) phonon driven.
\end{abstract}
\maketitle

\section{Introduction}

The emergence of unconventional superconductivity is often associated with the suppression of another electronic order \cite{Gegenwart2008,Stewart2011,Keimer2015,Fay1980}. In the parent compound of the Fe-based superconductors BaFe$_2$As$_2$, the leading instability is a uniaxial spin density wave (SDW) which breaks the rotational symmetry of the system and triggers a sizeable structural tetragonal to orthorhombic distortion \cite{Fernandes2014,Nandi2010,Rotter2008}, also known as nematic order. The critical enhancement of the nematic fluctuations upon the suppression of this transition and their coupling with electronic degrees of freedom largely shape the phase diagram of these materials. These fluctuations are believed to play an important role in the emergence of unconventional superconductivity \cite{Fernandes2014}, as the superconducting critical temperature $T_c $ is typically maximized in the region where the magneto-structural transition vanishes next to a quantum critical point \cite{Nandi2010,Boehmer2012,Hardy2010,Avci2012,Yamazaki2010}.

The superconductor BaNi$_2$As$_2$ \((T_c = \SI{0.7}{\kelvin})\) is a structural analogue to BaFe$_2$As$_2$ and crystallizes at ambient conditions in the ThCr$_2$Si$_2$ structure (\textit{I}4/\textit{mmm}) \cite{Pfisterer1980,Ronning2008}. Below \(\approx\SI{145}{\kelvin}\), instead of a SDW, an incommensurate charge density wave (I-CDW) order with wave vector \(q_\mathrm{I-CDW} = (0.28\ 0\ 0)_\mathrm{tet}\) or \((0\ 0.28\ 0)_\mathrm{tet}\) is formed, coincident with a small orthorhombic distortion to an \textit{Immm} symmetry \cite{Lee2019,Eckberg2020,Merz2021}. Two recent studies unveiled that the I-CDW is related to a complete phonon softening but is unconventional in nature, as no evidence of Fermi surface nesting or locally enhanced electron-phonon coupling was found \cite{Souliou2022, Song2023}. At a slightly lower temperature of \SI{139}{\kelvin} (upon cooling), the system undergoes a strong first-order structural transition to a triclinic (\textit{P}\(\overline{1}\)) phase. At the transition, the I-CDW is lost and a unidirectional commensurate CDW (C-CDW) with wavevector \(q_\mathrm{C-CDW} = (\pm1/3\ 0\ \mp1/3)_\mathrm{tet}\) or \((0\ \pm1/3\ \mp1/3)_\mathrm{tet}\) is formed \cite{Sefat2009,Lee2019}. Note that throughout the article, all Miller indices are given in the tetragonal notation irrespective of the actual crystal structure.  The subscript “tet” will generally be omitted for readability.

The C-CDW and triclinic phases can be suppressed by the application of hydrostatic pressure \cite{Lacmann2023} as well as by various substitutions, such as phosphorus \cite{Kudo2012,Meingast2022, Frachet2022}, strontium \cite{Eckberg2020}, cobalt \cite{Lee2019}, calcium \cite{Henssler2024} and copper \cite{kudo2017}, which additionally increase \(T_c\) up to \SI{3.5}{\kelvin}. Together with the recent observation of new pressure-induced CDW instabilities which are closely tied to structural changes \cite{Lacmann2023}, this suggests a strong interplay between the crystal structure and the CDWs.

Investigation of the phosphorus substituted system showed that in BaNi$_2$(As$_{1-\mathrm{x}}$P$_\mathrm{x}$) samples with a substitution level greater than \(x\approx0.075\) the C-CDW and triclinic phases are completely suppressed~\cite{Kudo2012, Yao2022, Meingast2022}. For higher concentrations (\(x\gtrapprox0.075\)), the in-plane orthorhombic distortion disappears \cite{Meingast2022}, even though the I-CDW is still observed by X-ray diffraction \cite{Yao2022}, and an out-of-plane lattice anomaly of unknown origin is reported \cite{Meingast2022}.
In addition, a soft-phonon contribution to the low-temperature specific heat and to an anomalous logarithmic divergence of the Grüneisen parameter upon cooling have been reported \cite{Meingast2022}.
Regardless of the presence or absence of the orthorhombic distortion, BaNi$_2$(As$_{1-\mathrm{x}}$P$_\mathrm{x}$)$_2$ is found to exhibit an unusually large splitting of the in-plane $E_g$ phonons which was explained through their coupling to \(B_{1g}\) nematic fluctuations \cite{Yao2022}.
These findings generally indicate a strong and unusual coupling between the electronic and lattice degrees of freedom in this material. However, there remain outstanding issues concerning the relationship between the C-CDW and I-CDW instabilities and the suppression of the orthorhombic distortion at high P-concentration calls for additional investigations on the evolution of the I-CDW instability as a function of phosphorus concentration.

In this paper, we address these issues by studying the evolution of the lattice dynamics anomalies associated with the formation of the I-CDW and C-CDW under phosphorus substitution. We performed a comprehensive thermal diffuse X-ray scattering (TDS) and (non-resonant) inelastic X-ray scattering (IXS) investigation on high quality phosphorus-substituted BaNi$_2$(As$_{1-\mathrm{x}}$P$_\mathrm{x}$)$_2$ \((x\lessapprox0.12)\) single crystals down to \SI{2.2}{\kelvin}. 
Our findings indicate that the structural instabilities (minute orthorhombic distortion or triclinic transition) in this system are intimately connected to the formation of a long-range 3D ordered CDW state. At high phosphorus contents  \((x \gtrapprox 0.075)\), such state never fully develops despite the full softening of a low lying optical mode at I-CDW and alongside a strong increase of the correlation lengths. This indicates that low-energy physics of these materials down to the lowest temperatures is governed by slow I-CDW fluctuations.

The result section of the paper is organized in four subsections. We first present the complex experimental reciprocal space TDS patterns above the I-CDW formation temperature for a representative phosphorus concentration \(x \approx 0.036\). These are analyzed in details on the basis of density functional perturbation theory (DFPT) phonon calculations that allow us to simulate the TDS pattern and reproduce most of the experimental features. A noticeable exception are sharp transverse peaks which emerge in the static I-CDW phase.
In the second section, we discuss the effects of temperature and P-concentration on the I-CDW. We report strong similarities in the formation of the I-CDW throughout the entire phosphorus substitution range  - investigated up to (\(x\lessapprox0.12\)) - with a modest lowering of the long-range ordering temperature and a broadening of the transition. In the third section, we investigate the phonon dispersion and we confirm that even at high phosphorus contents \(x\approx0.1\), and despite the suppression of the structural distortion, the I-CDW is soft-phonon driven. Finally in the fourth part, we negatively report on the search for lattice dynamics anomalies associated with the formation of the C-CDW.

\section{Methods}

\subsection{Crystal growth}
BaNi$_2$(As$_{1-\mathrm{x}}$P$_\mathrm{x}$)$_2$ single crystals were grown using the self-flux method. First, NiAs was synthesized by mixing Ni powder (Alfa Aesar 99.999\%) with As lumps (Alfa Aesar 99.9999\%) crushed into powder. The mixture was sealed under vacuum in a fused silica tube and annealed for \SI{20}{\hour} at \SI{730}{\celsius}.  For the final growth of BaNi$_2$(As$_{1-\mathrm{x}}$P$_\mathrm{x}$)$_2$ single crystals, a ratio of {Ba:NiAs:Ni:P = 1:4(1-x):4x:4x} precursor were placed in an alumina or glassy carbon crucible, which was sealed in an evacuated quartz ampule (i.e. \SI{1e-5}{\milli\bar}). All sample handling was performed inside an argon glove box (O$_2$ content \SI{0.7}{ppm}). The mixtures were heated to \SIrange[range-phrase={--}]{500}{700}{\celsius} for \SI{10}{\hour}, followed by slow heating to a temperature of \SIrange[range-phrase={--}]{1100}{1150}{\celsius}, soaked for \SI{5}{\hour}, and subsequently cooled to \SIrange[range-phrase={--}]{995}{950}{\celsius} at the rate of \SIrange{0.5}{1}{\celsius\per\hour}, depending on the phosphorus content used for the growth. At \SIrange[range-phrase={--}]{995}{950}{\celsius}, the furnace was canted to remove the excess flux, followed by furnace cooling. Plate-like single crystals with typical size \qtyproduct[product-units=power]{2 x 2 x 0.5}{\milli\metre} were easily removed from the remaining ingot. The crystals are brittle, with a shiny brass-yellow metallic luster. The chemical composition was determined with energy dispersive x-ray spectroscopy (EDX) using a COXEM EM-30AXN SEM-EDX scanning electron microscope with an Oxford Instruments Aztec EDX system. A full list of the sample investigated in this study is given in supplementary table~\ref{SM:tab:samples}, and the corresponding structural details have been reported in ref.~\cite{Meingast2022}.

\subsection{Thermal diffuse and inelastic X-ray scattering}
The thermal diffuse X-ray scattering and inelastic X-ray scattering experiments were performed at beamline ID28 of the European Synchrotron Radiation Facility (ESRF) \cite{Krisch2007,Girard2019}. For the IXS experiments, the incident X-ray beam energy was set to \SI{17.794}{\kilo\eV} using the \((9\ 9\ 9)\) silicon reflection with an energy resolution of \SI{3}{\milli\eV}. The beam was focused to a spot of \qtyproduct[product-units=power]{25 x 25}{\micro\meter} and the momentum resolution was set to \(\approx\SI{0.25}{\per\nano\meter}\) in the scattering plane and \(\approx\SI{0.75}{\per\nano\meter}\) perpendicular to it. The measurements on the \(x\approx0.1\) samples were performed with a closed-cycle cryostat with a final Joule-Thompson stage with Kapton windows and an inner aluminized Mylar heat shield (base temperature \(\approx\SI{2.2}{\kelvin}\)). The measurements on the \(x\approx0.037\) sample were performed with a Cryostream 700 Plus cooling system.

For the TDS measurements, the incident X-ray beam energy was set to \SI{17.794}{\kilo\eV} and the beam was focused to a spot of approximately \qtyproduct[product-units=power]{40 x 40}{\micro\meter}. The data were acquired with a Pilatus3 X 1M detector in shutterless mode while continuously rotating the sample and recording images while integrating over an angular range of \ang{0.25} and \SI{1}{\second}. The CrysAlis Pro software package~\cite{CrysalysPro} was used for determining the unit cell and sample orientation. The two- and three-dimensional reciprocal space reconstructions were created with a software developed at the beamline ID28. For all but the \(x\approx0.12\) sample low-temperature conditions were achieved using an Oxford Cryostream 700 Plus cooling system. For the \(x\approx0.12\) sample, a closed cycle cryostat (base temperature of \(\approx \SI{10}{\kelvin}\)) with Kapton windows was used. An internal screen and beamstop system on cryogenic bearings held in place by external magnets was employed to remove the resulting Kapton scattering. Details about the investigated samples can be found in the supplemental material (SM) \cite{SM}.

\subsection{Density functional perturbation theory and thermal diffuse scattering calculations} \label{Sec:DFPT}

Density functional perturbation theory (DFPT) calculations were performed, similar to those in the work from \citet{Souliou2022}. To stabilize the unstable phonon branch at the I-CDW position for the thermal diffuse scattering intensity calculations, the atomic positions were not minimized and a larger Gaussian smearing of \qtylist{0.1}{\eV} was used. The TDS scattering intensities were calculated to first order \cite{Wehinger2014} as implemented in an adapted version of the software ab2tds \cite{soft:ab2tds} with the DFPT calculations as input. To prevent imaginary or too low energies at the \(\Gamma\)-point leading to undefined TDS intensities, all phonon energies were capped at \qty{1e-4}{\milli\eV}. As it is generally challenging to include substitutions in DFPT, all calculations were performed for non-substituted BaNi$_2$As$_2$.
\section{Results}

\subsection{Thermal diffuse scattering above the triclinic transition}
\label{sec:TDS}

\begin{figure*}
    \includegraphics{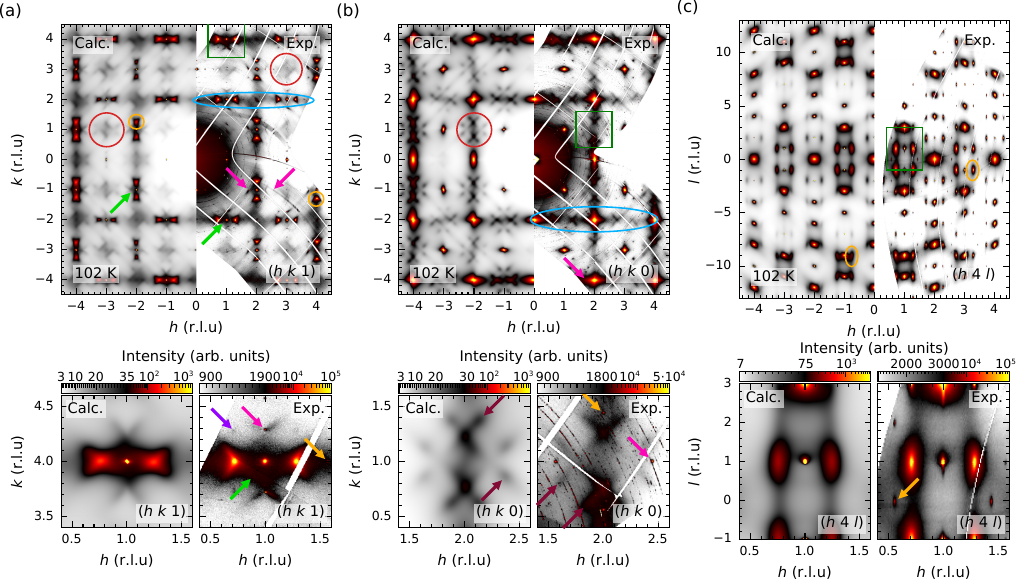}
    \caption{\label{Fig:DS:NewMaps} Thermal diffuse scattering calculations and comparison to experiment. (a)-(c) TDS calculations and experimental diffuse scattering maps in the (a) \((h\ k\ 1)\), (b) \((h\ k\ 0)\) and (c) \((h\ 4\ l)\)-planes at \qty{102}{\kelvin}. The experimental data are for the \(x\approx 0.036\) sample. Magnified views for the sections in the green rectangles are shown below for both the experimental and calculated TDS intensities. The following diffuse scattering features are highlighted: diffuse streaks between Bragg peaks (blue ellipse), rhomboids around allowed (green arrow) and forbidden Bragg reflections (ruby arrows), diffuse triangles at the I-CDW position (light orange ellipses) and their second-order reflection (light orange arrows), peaks in the longitudinal direction (pink arrow) and in the diagonal direction (purple arrow).}
\end{figure*}

We start our analysis of the diffuse scattering results by looking at  phosphorus-substituted BaNi$_2$(As$_{1-\mathrm{x}}$P$_\mathrm{x}$)$_2$ sample with \(x\approx 0.036\) at room temperature (above any structural transition) and in the I-CDW phase. 
Among these, the distinct triangular-shaped scattering around the I-CDW wavevector stands out as the most prominent and physically signifant signatures of the incipient I-CDW order. Specifically, the diffuse scattering maps in reciprocal space \cref{Fig:DS:NewMaps} reveal the following characteristic features:

\textbf{(i)} Diffuse triangular-shaped spots at the I-CDW position [e.g. \((1\pm0.28\ 4\ 1)\), highlighted by light orange ellipses], which are already visible at room temperature and evolve into sharp, Bragg-like peaks upon cooling. This features also exhibit second-order reflections (light orange arrows) at low temperature. Their distinct \(q-\) and temperature-dependence, along with their pronounced three-dimensional structure and absence in the  \((h\ k\ 0)\) plane, distinguish them from purely phononic features.

\textbf{(ii)} Diffuse streaks between Bragg peaks (highlighted by blue ellipses), visible in both \(h\) and \(k\) directions in the \((h\ k\ 0)\) and \((h\ k\ 1)\) planes. These features are well reproduced by TDS calculation and are consistent with a phononic origin.

\textbf{(iii)} Rhombus-shaped diffuse features around allowed Bragg reflections in e.g. the \((h\ k\ 1)\) plane (green arrows), with extended  “tails" along the diffuse lines. These structures are again well captured by the model and are akin to features observed in the parent compound\cite{Souliou2022}.

\textbf{(iv)} Similar Rhombus-shaped diffuse scattering is observed around forbidden Bragg peaks  [e.g., near \((2\ 1\ 0)\)], also accompanied by “tails" highlighted by ruby arrows. While TDS calculations generally agree with the experimental observations, the magnitude of the tail intensity is overestimated, confirmed by IXS measurements presented in the SM \cite{SM}. Notably, these locations correspond to the wave vector at which a second I-CDW emerges under pressure~\cite{Lacmann2023}.

\textbf{(v)} At low-temperatures, weak but sharp peaks appear in the longitudinal [e.g., \((1\ 4\pm0.28\ 1)\), pink arrows]
and diagonal directions [e.g., \((1\pm0.28\ 4\pm0.28\ 1)\), purple arrows] directions. These peaks, absent at room temperature, exhibit second-order reflections and are present even in the \((h\ k\ 0)\) plane, where the transverse diffuse features are not observed. Interestingly the diagonal peaks are not seen in the unsubstituted sample \cite{Souliou2022}.

\begin{figure*}
\includegraphics{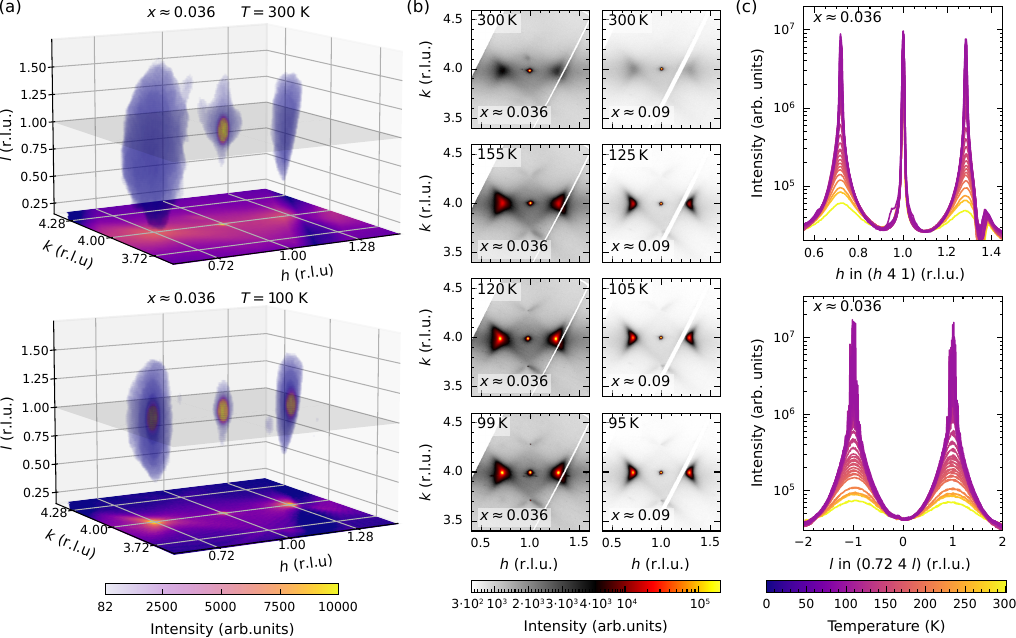}
\caption{\label{Fig:DS:Reconstructions} Three, two and one dimensional reciprocal space reconstructions around the I-CDW position. (a) Three dimensional reciprocal space reconstructions around the \((1\ 4\ 1)\) Bragg peak. The color and opacity indicate the intensity in each voxel. For visualization, the opacity of the colorbar is quadrupled. Parts of the peak at \((1.28\ 4\ 1\) are missing due to the empty space between the individual sensor modules. (b) \((h\ k\ 1)\) reciprocal space maps of the I-CDW diffuse scattering around the \((1\ 4\ 1)\) Bragg peak. Maps for \(x\approx0.036\) and \(\approx 0.09\) are shown. The maps represent the cut indicated by a gray plane in (a). (c) Linecuts in the \(h\) and \(l\) direction through the I-CDW position for the \(x\approx0.036\) sample at different temperatures.}
\end{figure*}

The evolution of the I-CDW diffuse feature is shown in \cref{Fig:DS:Reconstructions}. At room temperature, a triangular 
diffuse cloud is observed near  \((1.28\ 4\ 1)\), extending along the \(l-\)direction. This structure forms a three-dimensional shape which resembles a smoothed-out spherical wedge (‘kidney’) with a triangular cross-section, consistent with weaker interplanar correlations. As temperature decreases, this diffuse cloud sharpens and intensifies. A sharp, Bragg-like peak emerges at its center, signaling the formation of static I-CDW order. The diffuse component remains transverse in nature and is largely absent in the \((h\ k\ 0)\) plane. Model calculation accurately reproduce the shape and evolution of the diffuse component  (\cref{SM:Fig:DS:Calc3D}), but do not capture the sharp Bragg-like peaks, supporting the interpretation of a soft-phonon origin for the diffuse scattering and a static lattice distortion associated with the Bragg-like component.

Finally, we turn to the microscopic origin of the observed scattering features. Our model allows decomposition of the calculated TDS into contributions form different atomic species. The rhombus-shaped features and the diffuse scattering around the I-CDW wave vector arise predominantly from Ni and As displacements, while the enhanced tails near the forbidden Bragg reflections require contributions from all atomic species, including Ba. The combined TDS from all species is necessary to quantitatively capture the intensity distribution across the Brillouin zone. The atom-resolved 
TDS contributions are presented in \cref{SM:Fig:DS:Calc} of the SM \cite{SM}.

\subsection{Substitution dependence of the diffuse scattering}
\label{sec:TDS_dep}

\begin{figure}
    \includegraphics{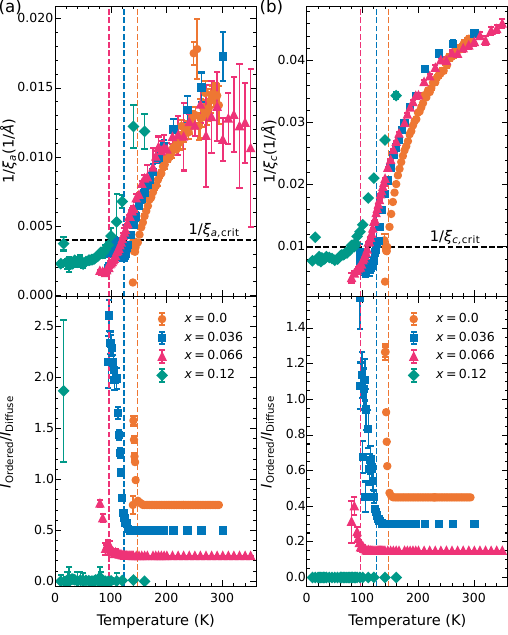}
    \caption{\label{fig:DS:CorLength} Temperature dependence of the inverse correlation length and the ratio of the ordered (pseudo-Voigt) and diffuse (Lorentzian) intensity in (a) the \(a\) and (b) the \(c\)-directions of the I-CDW for different substitution levels. Raw data for evaluation of the \(x=0\) is taken from \cite{Souliou2022}. The vertical lines indicate the transition temperatures extracted from the onset of the ordered intensity.}
\end{figure}

The diffuse scattering signal at the I-CDW position and the peaks in the longitudinal and diagonal direction also show a pronounced dependence with phosphorus concentration. 
Reciprocal space maps around the \((1\ 4\ 1)\) Bragg reflection, where these features are particularly intense, for samples with \(x\approx0.036\) and \(\approx 0.09\) over a range in temperatures are presented in \cref{Fig:DS:Reconstructions}(b). A complete substitution series and additional \((h\ 4\ l)\) maps can be found in \cref{SM:Fig:DS:HKMaps,SM:Fig:DS:HLMaps} in the supplementary material (SM) \cite{SM}.

In both compounds, the intense triangular diffuse scattering around the I-CDW positions becomes sharper and more intense upon cooling. At low temperatures, a sharp Bragg-like spot appears in the center of the diffuse scattering, and as this peak appears, second-order reflections and the longitudinal peaks also emerge [note that the diagonal features are also visible, but are extremely weak and hardly visible with the chosen contrast of \cref{Fig:DS:Reconstructions}(b)].  This behavior is observed for all the investigated substitution levels (up to \(x\approx0.12\)). The only noticeable difference among the various substitution levels is that the temperature at which the sharp peaks appear is progressively lowered and the peaks become increasingly smeared out.

The scattering signal at the transverse I-CDW wave vectors can be further quantified by making linecuts along the reciprocal \(h\) and \(k\) directions, as presented in \cref{Fig:DS:Reconstructions}(c) for the \(x\approx 0.036\) sample. Additional linecuts for other substitution contents and in the other directions (longitudinal and diagonal) display similar behavior and can be found in \cref{SM:Fig:DS:Cuts} of the SM \cite{SM}. 
The wavevector of the I-CDW can be directly obtained from the position of the superstructure peak in several Brillouin zones. While the temperature dependence remains small and is comparable for different substitution levels (\cref{SM:Fig:TDSWV}), the wavevector increases with higher phosphorous content. The trend is comparable to that observed under hydrostatic pressure but has a smaller magnitude \cite{Lacmann2023}. More details are presented in the SM \cite{SM}.

More quantitative analysis is obtained by fitting the linecuts through the I-CDW peaks using a single Lorentzian profile to further extract the intensity and full width at half maximum (FWHM) of the peak. The inverse correlation length can be calculated using $1/\xi = \mathrm{FWHM}/2$. The inverse correlation lengths in the \(a\)- and \(c\)-directions extracted from our analysis for all substitution levels are presented in \cref{fig:DS:CorLength}. For comparison, previously published data from the pristine sample \cite{Souliou2022} are also included in the figure. Additional fitting parameters are presented in \cref{SM:Fig:TDSFits} in the SM \cite{SM}. Notably, our analysis reveals that the correlation lengths at room temperature are only weakly dependent on the substitution levels, and are rather short, with \(\xi_a \approx \SI{60}{\angstrom}\) and \(\xi_c \approx \SI{22}{\angstrom}\). Similar to the pristine compound, both the in- and out-of-plane correlation lengths increase significantly upon cooling, rapidly reaching values as large as \(\xi_a \approx \SI{250}{\angstrom}\) and \(\xi_c \approx \SI{100}{\angstrom}\).

\begin{figure}
    \includegraphics{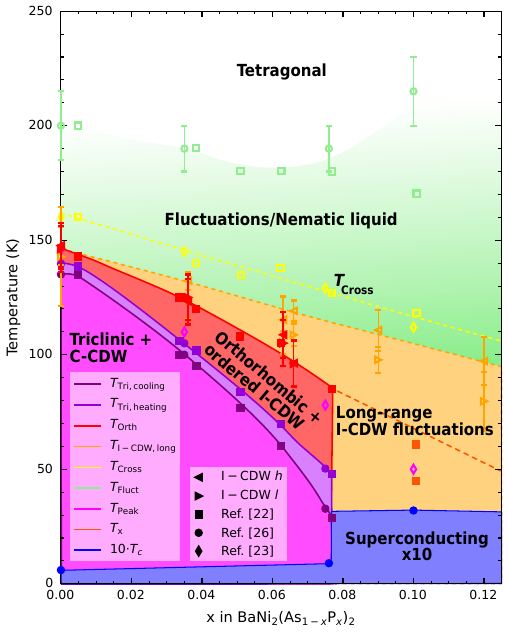}
    \caption{\label{fig:DS:PhaseDiagram} Phosphorus substitution phase diagram of BaNi$_2$(As$_{1-\mathrm{x}}$P$_\mathrm{x}$)$_2$. Included are real thermodynamic phase transitions determined from thermal-expansion and heat-capacity data in ref. \citet{Meingast2022}, the $T_{cross}$ temperature below which where the in-plane and out-of-plane thermal-expansion coefficients cross \citet{Meingast2022} and the elastoresistance signal onsets \citet{Frachet2022}, along with the onset of the nematic liquid obtained from Raman scattering \citet{Yao2022}. The  I-CDW and orthorhombic distortion presented in this diagram were determined in the present study.}

\end{figure}

A legitimate question arises: under what circumstances can one speak of true long-range I-CDW order? In principle, this should be associated with a thermodynamic phase transition, such as the subtle orthorhombic distortion detected via high-resolution thermal expansion measurements \cite{Merz2021, Meingast2022}. An independent criterion for identifying the onset of long-range order can be derived from the lineshape of the diffuse scattering features. As shown in Ref.~\cite{Souliou2022}, at sufficiently low temperatures, these profiles can no longer be accurately described by a single Lorentzian function. Instead, an additional Bragg-like superstructure reflection emerges, which is well-described by a resolution-limited pseudo-Voigt function whose shape and width are constrained by those of nearby main structural Bragg reflections. In the lower panels of \cref{fig:DS:CorLength}(a) and (b), we present the intensity ratio of the Bragg-like peak (pseudo-Voigt) to the diffuse scattering (Lorentzian), as obtained from our fitting procedure. The relatively sharp onset of a finite value of this ratio provides a well-defined criterion for determining the long-range ordering temperature from the x-ray data. The crossing point of a linear fit of the intensity ratio with 0 is used for the reliable definition of the transition temperature and corresponding error.
In line with previous works~\cite{Yao2022,Merz2021, Meingast2022}, it is found to continuously decrease as the phosphorus concentration increases. 
Interestingly, for low substitution contents we observe that the appearance of the Bragg-like contribution to the scattering intensity occurs  when the correlation length of the diffuse feature reaches a particular value. 
This “critical” correlation length  amounts to \(\xi_{\mathrm{a,crit}} \approx \SI{250}{\angstrom}\) in the \(a\)-direction and to \(\xi_{\mathrm{c,crit}} \approx \SI{100}{\angstrom}\) in the \(c\)-direction. 
Importantly, the \(\xi_{\mathrm{c,crit}}\) is always reached at a temperature slightly lower than that at which \(\xi_{\mathrm{a,crit}}\) is reached, but which agrees very well with that at which the thermodynamic investigation revealed the small structural distortion. Notably however, for the two highest substitution levels (\(x\approx0.09, 0.12\)) investigated, the Bragg-like signature of the long-range ordering of the I-CDW remains elusive even though both critical lengths are reached (measurements down to base temperature for the \(x\approx 0.12\) sample shows a saturation of the in- and out-of-plane correlation lengths well above the width of the Bragg peaks or experimental resolution). 
In these samples the C-CDW is completely suppressed and no orthorhombic distortion could be resolved in the \(ab\) plane~\cite{Meingast2022}. 
We have summarized the P-dependence of the transition temperatures obtained from x-ray scattering and thermodynamic measurements in the phase diagram in \cref{fig:DS:PhaseDiagram}.

Before turning to inelastic scattering investigation, we end this section noting that the weak elastic longitudinal and diagonal features as well as the second order reflections and signal at the I-CDW position in the \((h\ k\ 0)\)-plane can be observed also in the two highest substituted samples. These cannot easily be correlated with the appearance of the Bragg-like superstructure peaks \cite{SM}. Furthermore, the diagonal feature is only visible in the P-substituted samples. 
We note that the appearance of higher harmonic CDW reflections has been reported in a variety of systems encompassing  ${\mathrm{K}}_{2}$Pt${(\mathrm{CN})}_{4}$${\mathrm{Br}}_{0.3}$.3.2${\mathrm{D}}_{2}$O~\cite{Eagen1975} or LaAgSb$_2$~\cite{Song2003, Bosak2021}. They are often attributed to a squaring up of the sinusoidal charge density modulation, which however naturally favors odd harmonics against even ones. As argued by \citet{Song2003}, modulations of the lattice displacements on the other hand, can yield higher-harmonic satellite peaks at all orders even in presence of a pure sinusoidal charge modulation. To gain further insights, a full refinement of the atomic displacements induced by the CDW would be necessary, but lies well outside the scope of the present work. Further work would be required to assess their relationship to the small out-of-plane distortion reported at high P-content by high resolution dilatometry \cite{Meingast2022}.

\subsection{Inelastic X-Ray scattering of the I-CDW in phosphorus substituted samples (x=0.1)}

\begin{figure}
\includegraphics{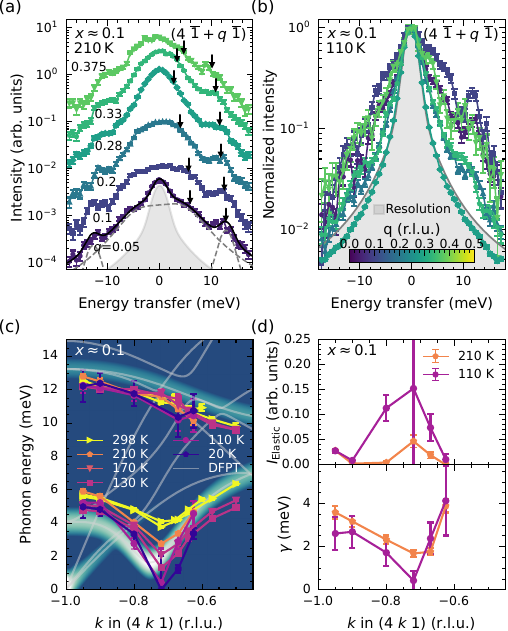}
\caption{\label{fig:IXS:ICDWDisp} Phonon dispersion around the I-CDW position. (a)-(b) IXS spectra along the \([0\ 1\ 0]\)-direction around the I-CDW wave vector at (a) \SI{210}{\kelvin} and (b) \SI{110}{\kelvin}. An example fit is shown in (a) and the fitted phonon positions are indicated by arrows. (c) Phonon dispersion along the \([0\ 1\ 0]\)-direction for different temperatures. In the background, the DFPT-calculated IXS intensity, broadened by a Gaussian with a FWHM of \qty{0.7}{\milli\eV}, is shown. The fitted dispersions for Samples 1 and 2 are shown, both of which were measured at \qty{298}{\kelvin} for reference. Hence, two dispersions are shown at this temperature. (d) Elastic intensity and phonon linewidth of the low-energy phonon for the dispersion in (c).}
\end{figure}

To get further insights regarding the mechanism yielding the formation of a long-range I-CDW order (or the lack thereof) in P-substituted BaNi$_2$As$_2$, we now turn to energy-resolved inelastic x-ray scattering. This follows previous investigations that have revealed a phonon softening associated with the I-CDW formation in the parent compound~\cite{Souliou2022, Song2023}. 

In \cref{fig:IXS:ICDWDisp} (a)-(b), we show characteristic IXS spectra for a \(x\approx0.1\) sample at \SI{210}{\kelvin} and \SI{110}{\kelvin}. The dispersions were measured along the \([0\ 1\ 0]\)-direction, with the \((4\ \overline{1}\ 1)\) as the \(\Gamma\)-point. Both the Stokes and anti-Stokes side are presented. In the spectra a resolution-limited elastic line and two dispersive phonons can be resolved, with the lowest energy showing a non-monotonic dispersion across the the I-CDW wave-vector at \((4\ \overline{0.72}\ 1)\) at 210 K. 
The IXS spectra are fitted with a resolution-limited elastic line and two damped harmonic oscillator functions (DHO), convoluted with the resolution functions. The resulting phonon dispersions at several temperatures are presented in \cref{fig:IXS:ICDWDisp} (c). While the higher energy phonon shows a normal dispersion and does not change significantly with temperature, a clear dip in the dispersion, or Kohn anomaly, of the lower phonon (which disperses from a zone center $E_g$ phonon previously investigated using Raman scattering~\cite{Yao2022}) can be seen already at room temperature. Upon cooling, the phonon softens further and can rapidly not be distinguished, with the experimental resolution, from the rapidly growing elastic line.
Notably, the intensity increase caused by the phonon softening and that caused by the increase of the elastic peak are very hard to disentangle with this energy resolution, and is hard to determine at which temperature the phonon becomes completely soft - even more so as the soft phonon is intrinsically very broad and close to being overdamped, as shown by the phonon linewidth in \cref{fig:IXS:ICDWDisp} (d).

\begin{figure}
\includegraphics{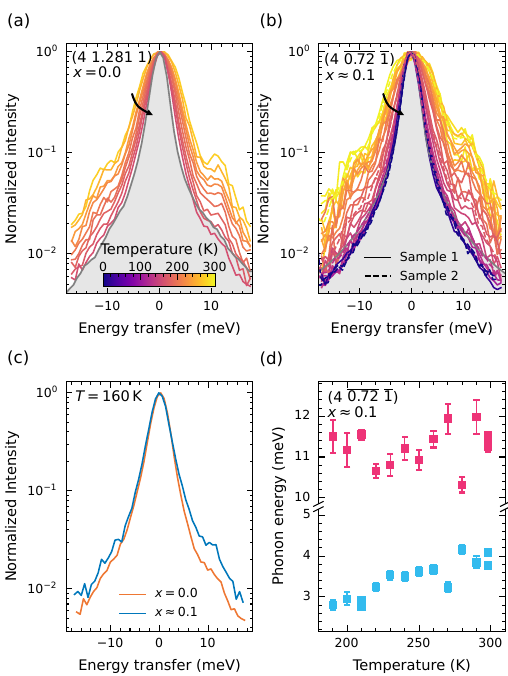}
\caption{\label{fig:IXS:ICDWTDep} Temperature dependence of the phonon instability at the I-CDW position. (a)-(b) Temperature dependence of the normalized IXS raw spectra at the I-CDW position for (a) the \(x=0.0\) and (b) the \(\approx 0.1\) samples. (c) Comparison of the normalized IXS spectra at \SI{160}{\kelvin} of \(x=0.0\) and \(\approx 0.1\). (d) Temperature dependence of the fitted phonon energy for \(x \approx 0.1\). IXS spectra for the \(x=0\) sample are reproduced from \cite{Souliou2022}.}
\end{figure}

This behavior is qualitatively comparable to that reported in parent BaNi$_2$As$_2$ in \cite{Souliou2022, Song2023}. In \cref{fig:IXS:ICDWTDep} (a)-(b), the temperature dependence of the IXS spectra at the I-CDW wave vector of the pristine and substituted sample are compared in greater detail. A one-to-one comparison of the lowest temperature possible is shown in \cref{fig:IXS:ICDWTDep} (c) and in \cref{SM:Fig:IXS:Comp} in the SM \cite{SM} for all temperatures. At each temperature, the overall profile of the IXS spectra is broader in the substituted sample, indicating that the softening is delayed with P-substitution, in line with the TDS results showing that the transition temperature is lowered. 

The fitting of the spectra of the \(x\approx0.1\) sample down to \SI{190}{\kelvin} shows that only the low-energy phonon softens, as the high-energy phonon does not change (see \ \cref{fig:IXS:ICDWTDep} (d)). 
At lower temperatures, the fits are omitted due to their unreliability, stemming from the low-energy and broad nature of the phonon mode, which cannot be reliably fitted without constraints as applied in the dispersion analysis.
Higher resolution investigation of the phonon softening \cite{Song2023} have concluded that the full softening occurs at a temperature slightly above that of the long-range ordering of the I-CDW. Although with the present \SI{3}{\milli\eV} resolution (and as in \cite{Souliou2022}) it is not possible to determine accurately the temperature at which the mode becomes completely soft, it is relatively clear that our results confirm the existence of a regime with a complete softening despite the absence of a long-range order (discussed in the previous section) at high P-concentration. 

We could not detect any phonon anomaly associated with the formation of the longitudinal satellites identified in the previous section, therefore confirming their elastic nature (see \cref{SM:Fig:IXS:ICDWOrthRaw} \cite{SM}).
To finish this section, we note that at the lowest investigated temperatures \(\SI{2.2}{\kelvin} < T_\mathrm{c}\) and \(\SI{3.75}{\kelvin} > T_\mathrm{c}\), and within the limitations described above, no competition between I-CDW and superconductivity can be identified. Indeed, no significant changes can be detected in the linewidth or phonon energy of the dispersion in the \([0\ 1\ 0]\)-direction (see \cref{SM:Fig:IXS:Tc} \cite{SM}).

\subsection{Inelastic X-Ray scattering around the I-CDW to C-CDW transition} 

\begin{figure}
\includegraphics{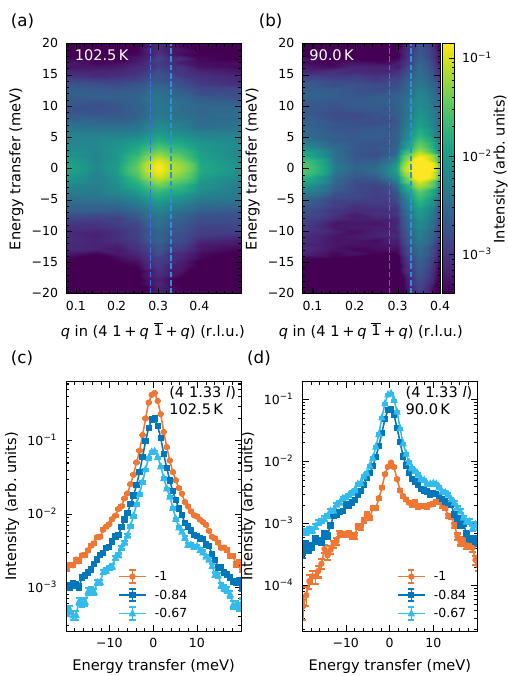}
\caption{\label{fig:IXS:CCDW} (a)-(b) Color map of the IXS spectra along the \([0\ 1\ 1]\)-direction around the C-CDW position for \(x\approx 0.037\). The dashed lines indicate the position closest to the I-CDW and the C-CDW, respectively. The measurements were taken at (a) \qty{102.5}{\kelvin}, just above the triclinic phase transition and (b) \SI{90}{\kelvin}, below the triclinic phase transition. (c)-(d) IXS scans along the \(l\)-direction through the C-CDW position at (c) \SI{102.5}{\kelvin}, just above the triclinic phase transition and (d) \SI{90}{\kelvin}, below the triclinic phase transition.}
\end{figure}

We finally address the nature of the transition from the I-CDW to the C-CDW in BaNi$_2$(As$_{1-\mathrm{x}}$P$_\mathrm{x}$)$_2$. As pointed out earlier, no precursor signatures of the C-CDW could be observed in the diffuse scattering maps, which strongly suggests the absence of a phonon softening related to the formation of the C-CDW.
To confirm this, IXS spectra along the \([0\ q\ q]\)-direction from the \((4\ 1\ \overline{1})\) Bragg reflection are shown in \cref{fig:IXS:CCDW} (a)-(b) (we remind here that the tetragonal notation is used). The measurements were performed on a \(x \approx 0.037\) sample, to enlarge the temperature range in which the I-CDW is present and have the transition to the C-CDW in an experimentally easily accessible temperature range (compared to a $\approx$ \SI{10}{\kelvin} range for the parent compound). The scans at \SI{102.5}{\kelvin}, just above the triclinic transition for this substitution level, show a pronounced sharpening towards \(q=0.28\), related to the softening of the low-lying phonon due to the I-CDW. This is confirmed by the scan along the \([0\ 0\ l]\)-direction around the C-CDW position (see \cref{fig:IXS:CCDW} (c)), indicating a stronger softening at \(l=-1\), as expected for the I-CDW.
Notably, at lower temperatures, no signature of a softening at the C-CDW position is visible. At \SI{90}{\kelvin}, just below the transition, the peak sharpens on approaching \(q=\frac{1}{3}\), indicating the static (elastic) C-CDW intensity. The spectrum at \(q=0.28\), however, broadens and again shows a normal mode. This is confirmed by the scans along \([0\ 0\ l]\) showing the centering at \(l=-\frac{2}{3}\), the C-CDW position.
\begin{figure}
\includegraphics{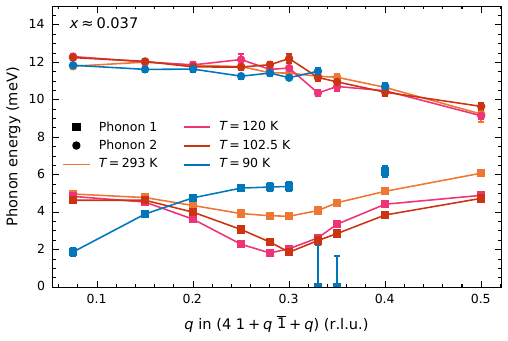}
\caption{\label{fig:IXS:CCDWDisp} Phonon dispersion along the \([0\ 1\ 1]\)-direction for different temperatures both above at 293, 120 and 102.5 K and below (at 90 K) the triclinic phase transition. The dispersions were measured on the \(x\approx 0.037\) sample. The two points at 0 energy for the 90 K data are the result of the fitting in presence of the strong elastic line at the C-CDW wave vector and should not be mistaken for a complete phonon softening.}
\end{figure}

The single spectra can be fitted, resulting in the phonon dispersion in \cref{fig:IXS:CCDWDisp}. At all temperatures, the upper phonon branch only marginally changes. The low-lying phonon, however, first shows a softening around \(q=0.28\) indicating the softening due to the I-CDW discussed above. At \(q=\frac{1}{3}\), no (additional) softening is visible above the phase transition. The changes in the measured dispersion at lower \(q\)-values reflect the structural changes occurring at the triclinic transition.
 The absence of a signature of phonon softening just above the transition temperature indicates that, in contrast to the I-CDW, the C-CDW transition is not driven by a distinct phonon instability.

\section{Discussion}

Our first principle and TDS calculations show that most features in the complex diffuse scattering maps can be attributed to phonons (the exceptions to this are the sharp longitudinal and diagonal peaks emerging at low temperatures). Our IXS and TDS data show a striking similarity in the formation of the I-CDW throughout the phosphorus substitution series, with an increase of the correlation length and a phonon softening at the I-CDW wave vector, even for \(x\approx0.1\), where the triclinic and C-CDW phase are completely suppressed. 
A 3D long-range ordering, characterized by the appearance of a resolution-limited Bragg-like feature on top of the diffuse scattering signal at the I-CDW wave vector, occurs when the correlation lengths in- and out-of-plane reach specific values which amount to \(\xi_{\mathrm{a,crit}} \approx \SI{250}{\angstrom}\) in the \(a\)-direction and to \(\xi_{\mathrm{c,crit}} \approx \SI{100}{\angstrom}\) in the \(c\)-direction.
Notably, this temperature exhibits significant anisotropy with increasing phosphorous content, occurring at a slightly higher temperature in the plane than along the \(c\)-axis.More precisely, for low phosphorus contents \((x\lesssim 0.075)\), the appearance temperature of the Bragg-like feature along the \(c\)-axis matches well that of the orthorhombic distortion detected by thermodynamic measurements \cite{Meingast2022}. For higher phosphorus contents \((x\gtrsim 0.075)\), however the in-plane correlation length of the I-CDW fluctuations grows larger than \(\xi_{\mathrm{a,crit}}\) but saturates along the c-axis down to base temperature to a value very close to \(\xi_{\mathrm{c,crit}}\). This correlates with the absence of realization of a fully coherent 3D long-range I-CDW ordering. Additionally, at these substitution levels, high-resolution thermal expansion measurements were not able to resolve an in-plane anisotropy, indicating the absence of the orthorhombic phase that forms at lower concentrations \cite{Meingast2022}. We can conclude from these observations that the thermodynamic phase transition associated with a minute orthorhombic distortion in the ab-plane is associated with the formation of a fully developed long-range 3D I-CDW ordering, which only seem to occur for P-concentrations less than approximately \((x = 0.075)\).

It is also worth noting that the suppression of this long-range order coincides with the disappearance of the C-CDW instability and the associated triclinic transition. Given that we have demonstrated this transition is not driven by a soft-phonon anomaly, a plausible interpretation is that it results from a lock-in of the 3D long-range I-CDW order. This aligns with time-resolved spectroscopy measurements, which suggest that the collective CDW excitations transition smoothly across the two CDW phases, indicating an intrinsic connection between them \cite{Pokharel2022}.

Additionally, we observe that near the temperature where the I-CDW correlation lengths saturate, thermodynamic measurements indicate that, at higher P-content, a different type of structural instability emerges along the c-axis, characterized by a slight expansion of the c-axis parameter below $\sim$ 80K~\cite{Meingast2022}. However our DS data do not reveal any structural distortion. Since DS is not ideally suited for precise structural refinement, further low-temperature structural investigations are needed.

Comparing the effect of P- and  Sr-substitution, in the strontium-substituted system, the I-CDW is completely suppressed, with only I-CDW fluctuations remaining at a substitution level just slightly higher than the point at which the triclinic phase with the C-CDW disappears ($x_{Sr}\sim 0.7$) \cite{chen2024}, bearing qualitative similarities with the present case. 
However, in the P-substituted system, the suppression of long-range I-CDW order appears to be accompanied by a significant lowering of both the temperature at which complete phonon softening occurs and that at which the (quasi-)elastic signal begins to grow. This is consistent with previous reports on the parent BaNi$_2$As$_2$, where the overdamping of phonons above the I-CDW ordering temperature and the emergence of slow, quasielastic I-CDW fluctuations were identified as key features governing the low-energy dynamics of the material\cite{Song2023}.

The in-plane correlation length associated with the I-CDW fluctuations are approximately three times shorter  \cite{chen2024} than that seen for $x\approx0.1$ P-sample. This indicates that a much larger I-CDW fluctuation regime upon P-substitution. 
Furthermore, while we cannot entirely rule out a competition between the I-CDW, or respectively I-CDW fluctuations, and superconductivity due to experimental limitations (we could only measure a temperature of \(T \sim 2/3\ T_c\) next to the sample, but were not able to measure the transition in situ), we found no evidence for such a competition - unlike in the superconducting cuprates \cite{Ghiringhelli2012,Chang2012,Vinograd2024} and other CDW materials \cite{Leroux2020,Stier2024}. Instead, the complete suppression of the C-CDW which correlates both in P- and Sr-substitution cases with a significant increase of the superconducting transition temperature appears more relevant to superconductivity in these systems than the presence of strong charge fluctuations. This supports BCS-type over exotic superconducting pairing mechanisms in these compounds.

Nonetheless, we emphasize that the presence of slow I-CDW fluctuations through the entire phase diagram of BaNi$_2$(As$_{1-\mathrm{x}}$P$_\mathrm{x}$)$_2$ likely relates to the observation of a large splitting of the doubly-degenerate \(\mathrm{E}_\mathrm{g}\) phonon modes at the \(\Gamma\)-point despite the absence of structural distortions, and that was interpreted as a consequence of their coupling to slow nematic fluctuations, out of which a nematic liquid behavior emerges \cite{Yao2022}. More generally, the fact that the I-CDW fluctuations are only marginally influenced by phosphorus substitution emphasizes their resilience, similar to the resilience of the I-CDW order found under hydrostatic pressure \cite{Lacmann2023}.

\section{Summary and outlook}

In summary, we have studied the formation of incommensurate and commensurate charge density waves for different samples of phosphorus substituted BaNi$_2$(As$_{1-\mathrm{x}}$P$_\mathrm{x}$)$_2$ \((0.036\lessapprox x \lessapprox 0.12)\) using X-ray thermal diffuse scattering and inelastic X-Ray scattering. Most of the complex thermal diffuse scattering patterns can be well described by first-principle phonon calculations. The in- and out-of-plane correlation lengths grow strongly upon cooling, yielding the formation of a 3D long-range I-CDW order for \((x \lessapprox 0.075)\), associated with a minute in-plane orthorhombic distortion and followed by a C-CDW/triclinic transition. 
The lack of phonon anomalies related to the formation of C-CDW suggests it might be resulting from a lock-in transition of the I-CDW concomitant to the large structural triclinic distortion which was previously found to enhance the occupation of the  $d_{xz}$ and $d_{yz}$ orbitals~\cite{Merz2021}, potentially impacting the charge modulation along the c-axis. For \((x \gtrapprox 0.075)\), we observe the full softening of a low lying optical mode at I-CDW alongside strong increase of the correlation lengths, but no formation of a long-range I-CDW order and associated structural distortions (and accordingly, the triclinic distortion and C-CDW are fully suppressed). 
This is indicative of a regime in which resilient, slow I-CDW fluctuations persist down to the lowest temperatures.

\section*{Data availability}

The raw IXS scans and the diffraction images from the TDS measurements are available from the Karlsruhe Institute of Technology repository KITOpen \cite{Data:KIT}.

\begin{acknowledgments}
We acknowledge the European Synchrotron Radiation Facility (ESRF) for provision of synchrotron radiation facilities under proposal numbers HC 4946, HC 5156 and HC 5340 and we would like to thank  A. Korshunov and L. Paolasini for assistance and support in using beamline ID28. We acknowledge the funding by the Deutsche Forschungsgemeinschaft (DFG; German Research Foundation) Project-ID 422213477-TRR 288 (Project B03) and support by the state of Baden-Württemberg through bwHPC. We thank T. Forrest for fruitful discussions.\\
\end{acknowledgments}

\bibliography{IXSDS2024}

\providecommand{\noopsort}[1]{}\providecommand{\singleletter}[1]{#1}%
\begin{thebibliography}{43}%
\makeatletter
\providecommand \@ifxundefined [1]{%
 \@ifx{#1\undefined}
}%
\providecommand \@ifnum [1]{%
 \ifnum #1\expandafter \@firstoftwo
 \else \expandafter \@secondoftwo
 \fi
}%
\providecommand \@ifx [1]{%
 \ifx #1\expandafter \@firstoftwo
 \else \expandafter \@secondoftwo
 \fi
}%
\providecommand \natexlab [1]{#1}%
\providecommand \enquote  [1]{``#1''}%
\providecommand \bibnamefont  [1]{#1}%
\providecommand \bibfnamefont [1]{#1}%
\providecommand \citenamefont [1]{#1}%
\providecommand \href@noop [0]{\@secondoftwo}%
\providecommand \href [0]{\begingroup \@sanitize@url \@href}%
\providecommand \@href[1]{\@@startlink{#1}\@@href}%
\providecommand \@@href[1]{\endgroup#1\@@endlink}%
\providecommand \@sanitize@url [0]{\catcode `\\12\catcode `\$12\catcode `\&12\catcode `\#12\catcode `\^12\catcode `\_12\catcode `\%12\relax}%
\providecommand \@@startlink[1]{}%
\providecommand \@@endlink[0]{}%
\providecommand \url  [0]{\begingroup\@sanitize@url \@url }%
\providecommand \@url [1]{\endgroup\@href {#1}{\urlprefix }}%
\providecommand \urlprefix  [0]{URL }%
\providecommand \Eprint [0]{\href }%
\providecommand \doibase [0]{https://doi.org/}%
\providecommand \selectlanguage [0]{\@gobble}%
\providecommand \bibinfo  [0]{\@secondoftwo}%
\providecommand \bibfield  [0]{\@secondoftwo}%
\providecommand \translation [1]{[#1]}%
\providecommand \BibitemOpen [0]{}%
\providecommand \bibitemStop [0]{}%
\providecommand \bibitemNoStop [0]{.\EOS\space}%
\providecommand \EOS [0]{\spacefactor3000\relax}%
\providecommand \BibitemShut  [1]{\csname bibitem#1\endcsname}%
\let\auto@bib@innerbib\@empty
\bibitem [{\citenamefont {Gegenwart}\ \emph {et~al.}(2008)\citenamefont {Gegenwart}, \citenamefont {Si},\ and\ \citenamefont {Steglich}}]{Gegenwart2008}%
  \BibitemOpen
  \bibfield  {author} {\bibinfo {author} {\bibfnamefont {P.}~\bibnamefont {Gegenwart}}, \bibinfo {author} {\bibfnamefont {Q.}~\bibnamefont {Si}},\ and\ \bibinfo {author} {\bibfnamefont {F.}~\bibnamefont {Steglich}},\ }\bibfield  {title} {\bibinfo {title} {Quantum criticality in heavy-fermion metals},\ }\href {https://doi.org/10.1038/nphys892} {\bibfield  {journal} {\bibinfo  {journal} {Nature Physics}\ }\textbf {\bibinfo {volume} {4}},\ \bibinfo {pages} {186} (\bibinfo {year} {2008})}\BibitemShut {NoStop}%
\bibitem [{\citenamefont {Stewart}(2011)}]{Stewart2011}%
  \BibitemOpen
  \bibfield  {author} {\bibinfo {author} {\bibfnamefont {G.~R.}\ \bibnamefont {Stewart}},\ }\bibfield  {title} {\bibinfo {title} {Superconductivity in iron compounds},\ }\href {https://doi.org/10.1103/revmodphys.83.1589} {\bibfield  {journal} {\bibinfo  {journal} {Reviews of Modern Physics}\ }\textbf {\bibinfo {volume} {83}},\ \bibinfo {pages} {1589} (\bibinfo {year} {2011})}\BibitemShut {NoStop}%
\bibitem [{\citenamefont {Keimer}\ \emph {et~al.}(2015)\citenamefont {Keimer}, \citenamefont {Kivelson}, \citenamefont {Norman}, \citenamefont {Uchida},\ and\ \citenamefont {Zaanen}}]{Keimer2015}%
  \BibitemOpen
  \bibfield  {author} {\bibinfo {author} {\bibfnamefont {B.}~\bibnamefont {Keimer}}, \bibinfo {author} {\bibfnamefont {S.~A.}\ \bibnamefont {Kivelson}}, \bibinfo {author} {\bibfnamefont {M.~R.}\ \bibnamefont {Norman}}, \bibinfo {author} {\bibfnamefont {S.}~\bibnamefont {Uchida}},\ and\ \bibinfo {author} {\bibfnamefont {J.}~\bibnamefont {Zaanen}},\ }\bibfield  {title} {\bibinfo {title} {From quantum matter to high-temperature superconductivity in copper oxides},\ }\href {https://doi.org/10.1038/nature14165} {\bibfield  {journal} {\bibinfo  {journal} {Nature}\ }\textbf {\bibinfo {volume} {518}},\ \bibinfo {pages} {179} (\bibinfo {year} {2015})}\BibitemShut {NoStop}%
\bibitem [{\citenamefont {Fay}\ and\ \citenamefont {Appel}(1980)}]{Fay1980}%
  \BibitemOpen
  \bibfield  {author} {\bibinfo {author} {\bibfnamefont {D.}~\bibnamefont {Fay}}\ and\ \bibinfo {author} {\bibfnamefont {J.}~\bibnamefont {Appel}},\ }\bibfield  {title} {\bibinfo {title} {Coexistence of p-state superconductivity and itinerant ferromagnetism},\ }\href {https://doi.org/10.1103/physrevb.22.3173} {\bibfield  {journal} {\bibinfo  {journal} {Physical Review B}\ }\textbf {\bibinfo {volume} {22}},\ \bibinfo {pages} {3173} (\bibinfo {year} {1980})}\BibitemShut {NoStop}%
\bibitem [{\citenamefont {Fernandes}\ \emph {et~al.}(2014)\citenamefont {Fernandes}, \citenamefont {Chubukov},\ and\ \citenamefont {Schmalian}}]{Fernandes2014}%
  \BibitemOpen
  \bibfield  {author} {\bibinfo {author} {\bibfnamefont {R.~M.}\ \bibnamefont {Fernandes}}, \bibinfo {author} {\bibfnamefont {A.~V.}\ \bibnamefont {Chubukov}},\ and\ \bibinfo {author} {\bibfnamefont {J.}~\bibnamefont {Schmalian}},\ }\bibfield  {title} {\bibinfo {title} {What drives nematic order in iron-based superconductors?},\ }\href {https://doi.org/10.1038/nphys2877} {\bibfield  {journal} {\bibinfo  {journal} {Nature Physics}\ }\textbf {\bibinfo {volume} {10}},\ \bibinfo {pages} {97} (\bibinfo {year} {2014})}\BibitemShut {NoStop}%
\bibitem [{\citenamefont {Nandi}\ \emph {et~al.}(2010)\citenamefont {Nandi}, \citenamefont {Kim}, \citenamefont {Kreyssig}, \citenamefont {Fernandes}, \citenamefont {Pratt}, \citenamefont {Thaler}, \citenamefont {Ni}, \citenamefont {Bud’ko}, \citenamefont {Canfield}, \citenamefont {Schmalian}, \citenamefont {McQueeney},\ and\ \citenamefont {Goldman}}]{Nandi2010}%
  \BibitemOpen
  \bibfield  {author} {\bibinfo {author} {\bibfnamefont {S.}~\bibnamefont {Nandi}}, \bibinfo {author} {\bibfnamefont {M.~G.}\ \bibnamefont {Kim}}, \bibinfo {author} {\bibfnamefont {A.}~\bibnamefont {Kreyssig}}, \bibinfo {author} {\bibfnamefont {R.~M.}\ \bibnamefont {Fernandes}}, \bibinfo {author} {\bibfnamefont {D.~K.}\ \bibnamefont {Pratt}}, \bibinfo {author} {\bibfnamefont {A.}~\bibnamefont {Thaler}}, \bibinfo {author} {\bibfnamefont {N.}~\bibnamefont {Ni}}, \bibinfo {author} {\bibfnamefont {S.~L.}\ \bibnamefont {Bud’ko}}, \bibinfo {author} {\bibfnamefont {P.~C.}\ \bibnamefont {Canfield}}, \bibinfo {author} {\bibfnamefont {J.}~\bibnamefont {Schmalian}}, \bibinfo {author} {\bibfnamefont {R.~J.}\ \bibnamefont {McQueeney}},\ and\ \bibinfo {author} {\bibfnamefont {A.~I.}\ \bibnamefont {Goldman}},\ }\bibfield  {title} {\bibinfo {title} {Anomalous suppression of the orthorhombic lattice distortion in superconducting {Ba(Fe\textsubscript{1-x}Co\textsubscript{x})\textsubscript{2}As\textsubscript{2}} single
  crystals},\ }\href {https://doi.org/10.1103/physrevlett.104.057006} {\bibfield  {journal} {\bibinfo  {journal} {Physical Review Letters}\ }\textbf {\bibinfo {volume} {104}},\ \bibinfo {pages} {057006} (\bibinfo {year} {2010})}\BibitemShut {NoStop}%
\bibitem [{\citenamefont {Rotter}\ \emph {et~al.}(2008)\citenamefont {Rotter}, \citenamefont {Tegel}, \citenamefont {Johrendt}, \citenamefont {Schellenberg}, \citenamefont {Hermes},\ and\ \citenamefont {Pöttgen}}]{Rotter2008}%
  \BibitemOpen
  \bibfield  {author} {\bibinfo {author} {\bibfnamefont {M.}~\bibnamefont {Rotter}}, \bibinfo {author} {\bibfnamefont {M.}~\bibnamefont {Tegel}}, \bibinfo {author} {\bibfnamefont {D.}~\bibnamefont {Johrendt}}, \bibinfo {author} {\bibfnamefont {I.}~\bibnamefont {Schellenberg}}, \bibinfo {author} {\bibfnamefont {W.}~\bibnamefont {Hermes}},\ and\ \bibinfo {author} {\bibfnamefont {R.}~\bibnamefont {Pöttgen}},\ }\bibfield  {title} {\bibinfo {title} {Spin-density-wave anomaly at 140 k in the ternary iron arsenide {BaFe\textsubscript{2}As\textsubscript{2}}},\ }\href {https://doi.org/10.1103/physrevb.78.020503} {\bibfield  {journal} {\bibinfo  {journal} {Physical Review B}\ }\textbf {\bibinfo {volume} {78}},\ \bibinfo {pages} {020503} (\bibinfo {year} {2008})}\BibitemShut {NoStop}%
\bibitem [{\citenamefont {Böhmer}\ \emph {et~al.}(2012)\citenamefont {Böhmer}, \citenamefont {Burger}, \citenamefont {Hardy}, \citenamefont {Wolf}, \citenamefont {Schweiss}, \citenamefont {Fromknecht}, \citenamefont {v.~Löhneysen}, \citenamefont {Meingast}, \citenamefont {Mak}, \citenamefont {Lortz}, \citenamefont {Kasahara}, \citenamefont {Terashima}, \citenamefont {Shibauchi},\ and\ \citenamefont {Matsuda}}]{Boehmer2012}%
  \BibitemOpen
  \bibfield  {author} {\bibinfo {author} {\bibfnamefont {A.~E.}\ \bibnamefont {Böhmer}}, \bibinfo {author} {\bibfnamefont {P.}~\bibnamefont {Burger}}, \bibinfo {author} {\bibfnamefont {F.}~\bibnamefont {Hardy}}, \bibinfo {author} {\bibfnamefont {T.}~\bibnamefont {Wolf}}, \bibinfo {author} {\bibfnamefont {P.}~\bibnamefont {Schweiss}}, \bibinfo {author} {\bibfnamefont {R.}~\bibnamefont {Fromknecht}}, \bibinfo {author} {\bibfnamefont {H.}~\bibnamefont {v.~Löhneysen}}, \bibinfo {author} {\bibfnamefont {C.}~\bibnamefont {Meingast}}, \bibinfo {author} {\bibfnamefont {H.~K.}\ \bibnamefont {Mak}}, \bibinfo {author} {\bibfnamefont {R.}~\bibnamefont {Lortz}}, \bibinfo {author} {\bibfnamefont {S.}~\bibnamefont {Kasahara}}, \bibinfo {author} {\bibfnamefont {T.}~\bibnamefont {Terashima}}, \bibinfo {author} {\bibfnamefont {T.}~\bibnamefont {Shibauchi}},\ and\ \bibinfo {author} {\bibfnamefont {Y.}~\bibnamefont {Matsuda}},\ }\bibfield  {title} {\bibinfo {title} {Thermodynamic phase diagram, phase competition, and uniaxial
  pressure effects in {BaFe\textsubscript{2}(As\textsubscript{1-x}P\textsubscript{x})\textsubscript{2}} studied by thermal expansion},\ }\href {https://doi.org/10.1103/physrevb.86.094521} {\bibfield  {journal} {\bibinfo  {journal} {Physical Review B}\ }\textbf {\bibinfo {volume} {86}},\ \bibinfo {pages} {094521} (\bibinfo {year} {2012})}\BibitemShut {NoStop}%
\bibitem [{\citenamefont {Hardy}\ \emph {et~al.}(2010)\citenamefont {Hardy}, \citenamefont {Burger}, \citenamefont {Wolf}, \citenamefont {Fisher}, \citenamefont {Schweiss}, \citenamefont {Adelmann}, \citenamefont {Heid}, \citenamefont {Fromknecht}, \citenamefont {Eder}, \citenamefont {Ernst}, \citenamefont {Löhneysen},\ and\ \citenamefont {Meingast}}]{Hardy2010}%
  \BibitemOpen
  \bibfield  {author} {\bibinfo {author} {\bibfnamefont {F.}~\bibnamefont {Hardy}}, \bibinfo {author} {\bibfnamefont {P.}~\bibnamefont {Burger}}, \bibinfo {author} {\bibfnamefont {T.}~\bibnamefont {Wolf}}, \bibinfo {author} {\bibfnamefont {R.~A.}\ \bibnamefont {Fisher}}, \bibinfo {author} {\bibfnamefont {P.}~\bibnamefont {Schweiss}}, \bibinfo {author} {\bibfnamefont {P.}~\bibnamefont {Adelmann}}, \bibinfo {author} {\bibfnamefont {R.}~\bibnamefont {Heid}}, \bibinfo {author} {\bibfnamefont {R.}~\bibnamefont {Fromknecht}}, \bibinfo {author} {\bibfnamefont {R.}~\bibnamefont {Eder}}, \bibinfo {author} {\bibfnamefont {D.}~\bibnamefont {Ernst}}, \bibinfo {author} {\bibfnamefont {H.~v.}\ \bibnamefont {Löhneysen}},\ and\ \bibinfo {author} {\bibfnamefont {C.}~\bibnamefont {Meingast}},\ }\bibfield  {title} {\bibinfo {title} {Doping evolution of superconducting gaps and electronic densities of states in {Ba(Fe\textsubscript{1-x}Co\textsubscript{x})\textsubscript{2}As\textsubscript{2}} iron pnictides},\ }\href
  {https://doi.org/10.1209/0295-5075/91/47008} {\bibfield  {journal} {\bibinfo  {journal} {EPL (Europhysics Letters)}\ }\textbf {\bibinfo {volume} {91}},\ \bibinfo {pages} {47008} (\bibinfo {year} {2010})}\BibitemShut {NoStop}%
\bibitem [{\citenamefont {Avci}\ \emph {et~al.}(2012)\citenamefont {Avci}, \citenamefont {Chmaissem}, \citenamefont {Chung}, \citenamefont {Rosenkranz}, \citenamefont {Goremychkin}, \citenamefont {Castellan}, \citenamefont {Todorov}, \citenamefont {Schlueter}, \citenamefont {Claus}, \citenamefont {Daoud-Aladine}, \citenamefont {Khalyavin}, \citenamefont {Kanatzidis},\ and\ \citenamefont {Osborn}}]{Avci2012}%
  \BibitemOpen
  \bibfield  {author} {\bibinfo {author} {\bibfnamefont {S.}~\bibnamefont {Avci}}, \bibinfo {author} {\bibfnamefont {O.}~\bibnamefont {Chmaissem}}, \bibinfo {author} {\bibfnamefont {D.~Y.}\ \bibnamefont {Chung}}, \bibinfo {author} {\bibfnamefont {S.}~\bibnamefont {Rosenkranz}}, \bibinfo {author} {\bibfnamefont {E.~A.}\ \bibnamefont {Goremychkin}}, \bibinfo {author} {\bibfnamefont {J.~P.}\ \bibnamefont {Castellan}}, \bibinfo {author} {\bibfnamefont {I.~S.}\ \bibnamefont {Todorov}}, \bibinfo {author} {\bibfnamefont {J.~A.}\ \bibnamefont {Schlueter}}, \bibinfo {author} {\bibfnamefont {H.}~\bibnamefont {Claus}}, \bibinfo {author} {\bibfnamefont {A.}~\bibnamefont {Daoud-Aladine}}, \bibinfo {author} {\bibfnamefont {D.~D.}\ \bibnamefont {Khalyavin}}, \bibinfo {author} {\bibfnamefont {M.~G.}\ \bibnamefont {Kanatzidis}},\ and\ \bibinfo {author} {\bibfnamefont {R.}~\bibnamefont {Osborn}},\ }\bibfield  {title} {\bibinfo {title} {Phase diagram of
  {Ba\textsubscript{1-x}K\textsubscript{x}Fe\textsubscript{2}As\textsubscript{2}}},\ }\href {https://doi.org/10.1103/physrevb.85.184507} {\bibfield  {journal} {\bibinfo  {journal} {Physical Review B}\ }\textbf {\bibinfo {volume} {85}},\ \bibinfo {pages} {184507} (\bibinfo {year} {2012})}\BibitemShut {NoStop}%
\bibitem [{\citenamefont {Yamazaki}\ \emph {et~al.}(2010)\citenamefont {Yamazaki}, \citenamefont {Takeshita}, \citenamefont {Kobayashi}, \citenamefont {Fukazawa}, \citenamefont {Kohori}, \citenamefont {Kihou}, \citenamefont {Lee}, \citenamefont {Kito}, \citenamefont {Iyo},\ and\ \citenamefont {Eisaki}}]{Yamazaki2010}%
  \BibitemOpen
  \bibfield  {author} {\bibinfo {author} {\bibfnamefont {T.}~\bibnamefont {Yamazaki}}, \bibinfo {author} {\bibfnamefont {N.}~\bibnamefont {Takeshita}}, \bibinfo {author} {\bibfnamefont {R.}~\bibnamefont {Kobayashi}}, \bibinfo {author} {\bibfnamefont {H.}~\bibnamefont {Fukazawa}}, \bibinfo {author} {\bibfnamefont {Y.}~\bibnamefont {Kohori}}, \bibinfo {author} {\bibfnamefont {K.}~\bibnamefont {Kihou}}, \bibinfo {author} {\bibfnamefont {C.-H.}\ \bibnamefont {Lee}}, \bibinfo {author} {\bibfnamefont {H.}~\bibnamefont {Kito}}, \bibinfo {author} {\bibfnamefont {A.}~\bibnamefont {Iyo}},\ and\ \bibinfo {author} {\bibfnamefont {H.}~\bibnamefont {Eisaki}},\ }\bibfield  {title} {\bibinfo {title} {Appearance of pressure-induced superconductivity in {BaFe\textsubscript{2}As\textsubscript{2}} under hydrostatic conditions and its extremely high sensitivity to uniaxial stress},\ }\href {https://doi.org/10.1103/physrevb.81.224511} {\bibfield  {journal} {\bibinfo  {journal} {Physical Review B}\ }\textbf {\bibinfo {volume}
  {81}},\ \bibinfo {pages} {224511} (\bibinfo {year} {2010})}\BibitemShut {NoStop}%
\bibitem [{\citenamefont {Pfisterer}\ and\ \citenamefont {Nagorsen}(1980)}]{Pfisterer1980}%
  \BibitemOpen
  \bibfield  {author} {\bibinfo {author} {\bibfnamefont {M.}~\bibnamefont {Pfisterer}}\ and\ \bibinfo {author} {\bibfnamefont {G.}~\bibnamefont {Nagorsen}},\ }\bibfield  {title} {\bibinfo {title} {Zur struktur ternärer Übergangsmetallarsenide},\ }\href {https://doi.org/doi:10.1515/znb-1980-0611} {\bibfield  {journal} {\bibinfo  {journal} {Zeitschrift für Naturforschung B}\ }\textbf {\bibinfo {volume} {35}},\ \bibinfo {pages} {703} (\bibinfo {year} {1980})}\BibitemShut {NoStop}%
\bibitem [{\citenamefont {Ronning}\ \emph {et~al.}(2008)\citenamefont {Ronning}, \citenamefont {Kurita}, \citenamefont {Bauer}, \citenamefont {Scott}, \citenamefont {Park}, \citenamefont {Klimczuk}, \citenamefont {Movshovich},\ and\ \citenamefont {Thompson}}]{Ronning2008}%
  \BibitemOpen
  \bibfield  {author} {\bibinfo {author} {\bibfnamefont {F.}~\bibnamefont {Ronning}}, \bibinfo {author} {\bibfnamefont {N.}~\bibnamefont {Kurita}}, \bibinfo {author} {\bibfnamefont {E.~D.}\ \bibnamefont {Bauer}}, \bibinfo {author} {\bibfnamefont {B.~L.}\ \bibnamefont {Scott}}, \bibinfo {author} {\bibfnamefont {T.}~\bibnamefont {Park}}, \bibinfo {author} {\bibfnamefont {T.}~\bibnamefont {Klimczuk}}, \bibinfo {author} {\bibfnamefont {R.}~\bibnamefont {Movshovich}},\ and\ \bibinfo {author} {\bibfnamefont {J.~D.}\ \bibnamefont {Thompson}},\ }\bibfield  {title} {\bibinfo {title} {The first order phase transition and superconductivity in {BaNi\textsubscript{2}As\textsubscript{2}} single crystals},\ }\href {https://doi.org/10.1088/0953-8984/20/34/342203} {\bibfield  {journal} {\bibinfo  {journal} {Journal of Physics: Condensed Matter}\ }\textbf {\bibinfo {volume} {20}},\ \bibinfo {pages} {342203} (\bibinfo {year} {2008})}\BibitemShut {NoStop}%
\bibitem [{\citenamefont {Lee}\ \emph {et~al.}(2019)\citenamefont {Lee}, \citenamefont {de~la Pe\~na}, \citenamefont {Sun}, \citenamefont {Mitrano}, \citenamefont {Fang}, \citenamefont {Jang}, \citenamefont {Lee}, \citenamefont {Eckberg}, \citenamefont {Campbell}, \citenamefont {Collini}, \citenamefont {Paglione}, \citenamefont {de~Groot},\ and\ \citenamefont {Abbamonte}}]{Lee2019}%
  \BibitemOpen
  \bibfield  {author} {\bibinfo {author} {\bibfnamefont {S.}~\bibnamefont {Lee}}, \bibinfo {author} {\bibfnamefont {G.}~\bibnamefont {de~la Pe\~na}}, \bibinfo {author} {\bibfnamefont {S.~X.-L.}\ \bibnamefont {Sun}}, \bibinfo {author} {\bibfnamefont {M.}~\bibnamefont {Mitrano}}, \bibinfo {author} {\bibfnamefont {Y.}~\bibnamefont {Fang}}, \bibinfo {author} {\bibfnamefont {H.}~\bibnamefont {Jang}}, \bibinfo {author} {\bibfnamefont {J.-S.}\ \bibnamefont {Lee}}, \bibinfo {author} {\bibfnamefont {C.}~\bibnamefont {Eckberg}}, \bibinfo {author} {\bibfnamefont {D.}~\bibnamefont {Campbell}}, \bibinfo {author} {\bibfnamefont {J.}~\bibnamefont {Collini}}, \bibinfo {author} {\bibfnamefont {J.}~\bibnamefont {Paglione}}, \bibinfo {author} {\bibfnamefont {F.~M.~F.}\ \bibnamefont {de~Groot}},\ and\ \bibinfo {author} {\bibfnamefont {P.}~\bibnamefont {Abbamonte}},\ }\bibfield  {title} {\bibinfo {title} {Unconventional charge density wave order in the pnictide superconductor
  {$\mathrm{Ba}({\mathrm{Ni}}_{1\ensuremath{-}x}{\mathrm{Co}}_{x}{)}_{2}{\mathrm{As}}_{2}$}},\ }\href {https://doi.org/10.1103/PhysRevLett.122.147601} {\bibfield  {journal} {\bibinfo  {journal} {Phys. Rev. Lett.}\ }\textbf {\bibinfo {volume} {122}},\ \bibinfo {pages} {147601} (\bibinfo {year} {2019})}\BibitemShut {NoStop}%
\bibitem [{\citenamefont {Eckberg}\ \emph {et~al.}(2020)\citenamefont {Eckberg}, \citenamefont {Campbell}, \citenamefont {Metz}, \citenamefont {Collini}, \citenamefont {Hodovanets}, \citenamefont {Drye}, \citenamefont {Zavalij}, \citenamefont {Christensen}, \citenamefont {Fernandes}, \citenamefont {Lee}, \citenamefont {Abbamonte}, \citenamefont {Lynn},\ and\ \citenamefont {Paglione}}]{Eckberg2020}%
  \BibitemOpen
  \bibfield  {author} {\bibinfo {author} {\bibfnamefont {C.}~\bibnamefont {Eckberg}}, \bibinfo {author} {\bibfnamefont {D.~J.}\ \bibnamefont {Campbell}}, \bibinfo {author} {\bibfnamefont {T.}~\bibnamefont {Metz}}, \bibinfo {author} {\bibfnamefont {J.}~\bibnamefont {Collini}}, \bibinfo {author} {\bibfnamefont {H.}~\bibnamefont {Hodovanets}}, \bibinfo {author} {\bibfnamefont {T.}~\bibnamefont {Drye}}, \bibinfo {author} {\bibfnamefont {P.}~\bibnamefont {Zavalij}}, \bibinfo {author} {\bibfnamefont {M.~H.}\ \bibnamefont {Christensen}}, \bibinfo {author} {\bibfnamefont {R.~M.}\ \bibnamefont {Fernandes}}, \bibinfo {author} {\bibfnamefont {S.}~\bibnamefont {Lee}}, \bibinfo {author} {\bibfnamefont {P.}~\bibnamefont {Abbamonte}}, \bibinfo {author} {\bibfnamefont {J.~W.}\ \bibnamefont {Lynn}},\ and\ \bibinfo {author} {\bibfnamefont {J.}~\bibnamefont {Paglione}},\ }\bibfield  {title} {\bibinfo {title} {Sixfold enhancement of superconductivity in a tunable electronic nematic system},\ }\href
  {https://doi.org/10.1038/s41567-019-0736-9} {\bibfield  {journal} {\bibinfo  {journal} {Nature Physics}\ }\textbf {\bibinfo {volume} {16}},\ \bibinfo {pages} {346} (\bibinfo {year} {2020})}\BibitemShut {NoStop}%
\bibitem [{\citenamefont {Merz}\ \emph {et~al.}(2021)\citenamefont {Merz}, \citenamefont {Wang}, \citenamefont {Wolf}, \citenamefont {Nagel}, \citenamefont {Meingast},\ and\ \citenamefont {Schuppler}}]{Merz2021}%
  \BibitemOpen
  \bibfield  {author} {\bibinfo {author} {\bibfnamefont {M.}~\bibnamefont {Merz}}, \bibinfo {author} {\bibfnamefont {L.}~\bibnamefont {Wang}}, \bibinfo {author} {\bibfnamefont {T.}~\bibnamefont {Wolf}}, \bibinfo {author} {\bibfnamefont {P.}~\bibnamefont {Nagel}}, \bibinfo {author} {\bibfnamefont {C.}~\bibnamefont {Meingast}},\ and\ \bibinfo {author} {\bibfnamefont {S.}~\bibnamefont {Schuppler}},\ }\bibfield  {title} {\bibinfo {title} {Rotational symmetry breaking at the incommensurate charge-density-wave transition in $\mathrm{Ba}{(\mathrm{Ni},\mathrm{Co})}_{2}{(\mathrm{As},\mathrm{P})}_{2}$: Possible nematic phase induced by charge/orbital fluctuations},\ }\href {https://doi.org/10.1103/PhysRevB.104.184509} {\bibfield  {journal} {\bibinfo  {journal} {Phys. Rev. B}\ }\textbf {\bibinfo {volume} {104}},\ \bibinfo {pages} {184509} (\bibinfo {year} {2021})}\BibitemShut {NoStop}%
\bibitem [{\citenamefont {Souliou}\ \emph {et~al.}(2022)\citenamefont {Souliou}, \citenamefont {Lacmann}, \citenamefont {Heid}, \citenamefont {Meingast}, \citenamefont {Frachet}, \citenamefont {Paolasini}, \citenamefont {Haghighirad}, \citenamefont {Merz}, \citenamefont {Bosak},\ and\ \citenamefont {Le~Tacon}}]{Souliou2022}%
  \BibitemOpen
  \bibfield  {author} {\bibinfo {author} {\bibfnamefont {S.~M.}\ \bibnamefont {Souliou}}, \bibinfo {author} {\bibfnamefont {T.}~\bibnamefont {Lacmann}}, \bibinfo {author} {\bibfnamefont {R.}~\bibnamefont {Heid}}, \bibinfo {author} {\bibfnamefont {C.}~\bibnamefont {Meingast}}, \bibinfo {author} {\bibfnamefont {M.}~\bibnamefont {Frachet}}, \bibinfo {author} {\bibfnamefont {L.}~\bibnamefont {Paolasini}}, \bibinfo {author} {\bibfnamefont {A.-A.}\ \bibnamefont {Haghighirad}}, \bibinfo {author} {\bibfnamefont {M.}~\bibnamefont {Merz}}, \bibinfo {author} {\bibfnamefont {A.}~\bibnamefont {Bosak}},\ and\ \bibinfo {author} {\bibfnamefont {M.}~\bibnamefont {Le~Tacon}},\ }\bibfield  {title} {\bibinfo {title} {Soft-phonon and charge-density-wave formation in nematic {${\mathrm{BaNi}}_{2}{\mathrm{As}}_{2}$}},\ }\href {https://doi.org/10.1103/PhysRevLett.129.247602} {\bibfield  {journal} {\bibinfo  {journal} {Phys. Rev. Lett.}\ }\textbf {\bibinfo {volume} {129}},\ \bibinfo {pages} {247602} (\bibinfo {year}
  {2022})}\BibitemShut {NoStop}%
\bibitem [{\citenamefont {Song}\ \emph {et~al.}(2023)\citenamefont {Song}, \citenamefont {Wu}, \citenamefont {Chen}, \citenamefont {He}, \citenamefont {Uchiyama}, \citenamefont {Li}, \citenamefont {Cao}, \citenamefont {Guo}, \citenamefont {Cao},\ and\ \citenamefont {Birgeneau}}]{Song2023}%
  \BibitemOpen
  \bibfield  {author} {\bibinfo {author} {\bibfnamefont {Y.}~\bibnamefont {Song}}, \bibinfo {author} {\bibfnamefont {S.}~\bibnamefont {Wu}}, \bibinfo {author} {\bibfnamefont {X.}~\bibnamefont {Chen}}, \bibinfo {author} {\bibfnamefont {Y.}~\bibnamefont {He}}, \bibinfo {author} {\bibfnamefont {H.}~\bibnamefont {Uchiyama}}, \bibinfo {author} {\bibfnamefont {B.}~\bibnamefont {Li}}, \bibinfo {author} {\bibfnamefont {S.}~\bibnamefont {Cao}}, \bibinfo {author} {\bibfnamefont {J.}~\bibnamefont {Guo}}, \bibinfo {author} {\bibfnamefont {G.}~\bibnamefont {Cao}},\ and\ \bibinfo {author} {\bibfnamefont {R.}~\bibnamefont {Birgeneau}},\ }\bibfield  {title} {\bibinfo {title} {Phonon softening and slowing-down of charge density wave fluctuations in {${\mathrm{BaNi}}_{2}{\mathrm{As}}_{2}$}},\ }\href {https://doi.org/10.1103/PhysRevB.107.L041113} {\bibfield  {journal} {\bibinfo  {journal} {Phys. Rev. B}\ }\textbf {\bibinfo {volume} {107}},\ \bibinfo {pages} {L041113} (\bibinfo {year} {2023})}\BibitemShut {NoStop}%
\bibitem [{\citenamefont {Sefat}\ \emph {et~al.}(2009)\citenamefont {Sefat}, \citenamefont {McGuire}, \citenamefont {Jin}, \citenamefont {Sales}, \citenamefont {Mandrus}, \citenamefont {Ronning}, \citenamefont {Bauer},\ and\ \citenamefont {Mozharivskyj}}]{Sefat2009}%
  \BibitemOpen
  \bibfield  {author} {\bibinfo {author} {\bibfnamefont {A.~S.}\ \bibnamefont {Sefat}}, \bibinfo {author} {\bibfnamefont {M.~A.}\ \bibnamefont {McGuire}}, \bibinfo {author} {\bibfnamefont {R.}~\bibnamefont {Jin}}, \bibinfo {author} {\bibfnamefont {B.~C.}\ \bibnamefont {Sales}}, \bibinfo {author} {\bibfnamefont {D.}~\bibnamefont {Mandrus}}, \bibinfo {author} {\bibfnamefont {F.}~\bibnamefont {Ronning}}, \bibinfo {author} {\bibfnamefont {E.~D.}\ \bibnamefont {Bauer}},\ and\ \bibinfo {author} {\bibfnamefont {Y.}~\bibnamefont {Mozharivskyj}},\ }\bibfield  {title} {\bibinfo {title} {Structure and anisotropic properties of {${\text{BaFe}}_{2\ensuremath{-}x}{\text{Ni}}_{x}{\text{As}}_{2}$} ($x=0$, 1, and 2) single crystals},\ }\href {https://doi.org/10.1103/PhysRevB.79.094508} {\bibfield  {journal} {\bibinfo  {journal} {Phys. Rev. B}\ }\textbf {\bibinfo {volume} {79}},\ \bibinfo {pages} {094508} (\bibinfo {year} {2009})}\BibitemShut {NoStop}%
\bibitem [{\citenamefont {Lacmann}\ \emph {et~al.}(2023)\citenamefont {Lacmann}, \citenamefont {Haghighirad}, \citenamefont {Souliou}, \citenamefont {Merz}, \citenamefont {Garbarino}, \citenamefont {Glazyrin}, \citenamefont {Heid},\ and\ \citenamefont {Le~Tacon}}]{Lacmann2023}%
  \BibitemOpen
  \bibfield  {author} {\bibinfo {author} {\bibfnamefont {T.}~\bibnamefont {Lacmann}}, \bibinfo {author} {\bibfnamefont {A.-A.}\ \bibnamefont {Haghighirad}}, \bibinfo {author} {\bibfnamefont {S.-M.}\ \bibnamefont {Souliou}}, \bibinfo {author} {\bibfnamefont {M.}~\bibnamefont {Merz}}, \bibinfo {author} {\bibfnamefont {G.}~\bibnamefont {Garbarino}}, \bibinfo {author} {\bibfnamefont {K.}~\bibnamefont {Glazyrin}}, \bibinfo {author} {\bibfnamefont {R.}~\bibnamefont {Heid}},\ and\ \bibinfo {author} {\bibfnamefont {M.}~\bibnamefont {Le~Tacon}},\ }\bibfield  {title} {\bibinfo {title} {High-pressure phase diagram of {${\mathrm{BaNi}}_{2}{\mathrm{As}}_{2}$}: Unconventional charge density waves and structural phase transitions},\ }\href {https://doi.org/10.1103/PhysRevB.108.224115} {\bibfield  {journal} {\bibinfo  {journal} {Phys. Rev. B}\ }\textbf {\bibinfo {volume} {108}},\ \bibinfo {pages} {224115} (\bibinfo {year} {2023})}\BibitemShut {NoStop}%
\bibitem [{\citenamefont {Kudo}\ \emph {et~al.}(2012)\citenamefont {Kudo}, \citenamefont {Takasuga}, \citenamefont {Okamoto}, \citenamefont {Hiroi},\ and\ \citenamefont {Nohara}}]{Kudo2012}%
  \BibitemOpen
  \bibfield  {author} {\bibinfo {author} {\bibfnamefont {K.}~\bibnamefont {Kudo}}, \bibinfo {author} {\bibfnamefont {M.}~\bibnamefont {Takasuga}}, \bibinfo {author} {\bibfnamefont {Y.}~\bibnamefont {Okamoto}}, \bibinfo {author} {\bibfnamefont {Z.}~\bibnamefont {Hiroi}},\ and\ \bibinfo {author} {\bibfnamefont {M.}~\bibnamefont {Nohara}},\ }\bibfield  {title} {\bibinfo {title} {Giant phonon softening and enhancement of superconductivity by phosphorus doping of {${\mathrm{BaNi}}_{2}{\mathrm{As}}_{2}$}},\ }\href {https://doi.org/10.1103/PhysRevLett.109.097002} {\bibfield  {journal} {\bibinfo  {journal} {Phys. Rev. Lett.}\ }\textbf {\bibinfo {volume} {109}},\ \bibinfo {pages} {097002} (\bibinfo {year} {2012})}\BibitemShut {NoStop}%
\bibitem [{\citenamefont {Meingast}\ \emph {et~al.}(2022)\citenamefont {Meingast}, \citenamefont {Shukla}, \citenamefont {Wang}, \citenamefont {Heid}, \citenamefont {Hardy}, \citenamefont {Frachet}, \citenamefont {Willa}, \citenamefont {Lacmann}, \citenamefont {Le~Tacon}, \citenamefont {Merz}, \citenamefont {Haghighirad},\ and\ \citenamefont {Wolf}}]{Meingast2022}%
  \BibitemOpen
  \bibfield  {author} {\bibinfo {author} {\bibfnamefont {C.}~\bibnamefont {Meingast}}, \bibinfo {author} {\bibfnamefont {A.}~\bibnamefont {Shukla}}, \bibinfo {author} {\bibfnamefont {L.}~\bibnamefont {Wang}}, \bibinfo {author} {\bibfnamefont {R.}~\bibnamefont {Heid}}, \bibinfo {author} {\bibfnamefont {F.}~\bibnamefont {Hardy}}, \bibinfo {author} {\bibfnamefont {M.}~\bibnamefont {Frachet}}, \bibinfo {author} {\bibfnamefont {K.}~\bibnamefont {Willa}}, \bibinfo {author} {\bibfnamefont {T.}~\bibnamefont {Lacmann}}, \bibinfo {author} {\bibfnamefont {M.}~\bibnamefont {Le~Tacon}}, \bibinfo {author} {\bibfnamefont {M.}~\bibnamefont {Merz}}, \bibinfo {author} {\bibfnamefont {A.-A.}\ \bibnamefont {Haghighirad}},\ and\ \bibinfo {author} {\bibfnamefont {T.}~\bibnamefont {Wolf}},\ }\bibfield  {title} {\bibinfo {title} {Charge density wave transitions, soft phonon, and possible electronic nematicity in {${\mathrm{BaNi}}_{2}({\mathrm{As}}_{1\ensuremath{-}x}{\mathrm{P}}_{x}{)}_{2}$}},\ }\href
  {https://doi.org/10.1103/PhysRevB.106.144507} {\bibfield  {journal} {\bibinfo  {journal} {Phys. Rev. B}\ }\textbf {\bibinfo {volume} {106}},\ \bibinfo {pages} {144507} (\bibinfo {year} {2022})}\BibitemShut {NoStop}%
\bibitem [{\citenamefont {Frachet}\ \emph {et~al.}(2022)\citenamefont {Frachet}, \citenamefont {Wiecki}, \citenamefont {Lacmann}, \citenamefont {Souliou}, \citenamefont {Willa}, \citenamefont {Meingast}, \citenamefont {Merz}, \citenamefont {Haghighirad}, \citenamefont {Le~Tacon},\ and\ \citenamefont {B{\"o}hmer}}]{Frachet2022}%
  \BibitemOpen
  \bibfield  {author} {\bibinfo {author} {\bibfnamefont {M.}~\bibnamefont {Frachet}}, \bibinfo {author} {\bibfnamefont {P.}~\bibnamefont {Wiecki}}, \bibinfo {author} {\bibfnamefont {T.}~\bibnamefont {Lacmann}}, \bibinfo {author} {\bibfnamefont {S.~M.}\ \bibnamefont {Souliou}}, \bibinfo {author} {\bibfnamefont {K.}~\bibnamefont {Willa}}, \bibinfo {author} {\bibfnamefont {C.}~\bibnamefont {Meingast}}, \bibinfo {author} {\bibfnamefont {M.}~\bibnamefont {Merz}}, \bibinfo {author} {\bibfnamefont {A.-A.}\ \bibnamefont {Haghighirad}}, \bibinfo {author} {\bibfnamefont {M.}~\bibnamefont {Le~Tacon}},\ and\ \bibinfo {author} {\bibfnamefont {A.~E.}\ \bibnamefont {B{\"o}hmer}},\ }\bibfield  {title} {\bibinfo {title} {Elastoresistivity in the incommensurate charge density wave phase of {BaNi\textsubscript{2}(As\textsubscript{1-x}P\textsubscript{x})\textsubscript{2}}},\ }\href {https://doi.org/10.1038/s41535-022-00525-8} {\bibfield  {journal} {\bibinfo  {journal} {npj Quantum Materials}\ }\textbf {\bibinfo {volume} {7}},\
  \bibinfo {pages} {115} (\bibinfo {year} {2022})}\BibitemShut {NoStop}%
\bibitem [{\citenamefont {Henssler}\ \emph {et~al.}(2025)\citenamefont {Henssler}, \citenamefont {Willa}, \citenamefont {Frachet}, \citenamefont {Lacmann}, \citenamefont {Chaney}, \citenamefont {Heid}, \citenamefont {Merz}, \citenamefont {Haghighirad},\ and\ \citenamefont {Le~Tacon}}]{Henssler2024}%
  \BibitemOpen
  \bibfield  {author} {\bibinfo {author} {\bibfnamefont {F.}~\bibnamefont {Henssler}}, \bibinfo {author} {\bibfnamefont {K.}~\bibnamefont {Willa}}, \bibinfo {author} {\bibfnamefont {M.}~\bibnamefont {Frachet}}, \bibinfo {author} {\bibfnamefont {T.}~\bibnamefont {Lacmann}}, \bibinfo {author} {\bibfnamefont {D.~A.}\ \bibnamefont {Chaney}}, \bibinfo {author} {\bibfnamefont {R.}~\bibnamefont {Heid}}, \bibinfo {author} {\bibfnamefont {M.}~\bibnamefont {Merz}}, \bibinfo {author} {\bibfnamefont {A.-A.}\ \bibnamefont {Haghighirad}},\ and\ \bibinfo {author} {\bibfnamefont {M.}~\bibnamefont {Le~Tacon}},\ }\bibfield  {title} {\bibinfo {title} {Impact of {Ca} substitution on competing orders in superconducting {${\mathrm{BaNi}}_{2}{\mathrm{As}}_{2}$}},\ }\href {https://doi.org/10.1103/PhysRevMaterials.9.044801} {\bibfield  {journal} {\bibinfo  {journal} {Phys. Rev. Mater.}\ }\textbf {\bibinfo {volume} {9}},\ \bibinfo {pages} {044801} (\bibinfo {year} {2025})}\BibitemShut {NoStop}%
\bibitem [{\citenamefont {Kudo}\ \emph {et~al.}(2017)\citenamefont {Kudo}, \citenamefont {Takasuga},\ and\ \citenamefont {Nohara}}]{kudo2017}%
  \BibitemOpen
  \bibfield  {author} {\bibinfo {author} {\bibfnamefont {K.}~\bibnamefont {Kudo}}, \bibinfo {author} {\bibfnamefont {M.}~\bibnamefont {Takasuga}},\ and\ \bibinfo {author} {\bibfnamefont {M.}~\bibnamefont {Nohara}},\ }\href@noop {} {\bibinfo {title} {Copper doping of {BaNi$_{2}$As$_{2}$}: Giant phonon softening and superconductivity enhancement}} (\bibinfo {year} {2017}),\ \Eprint {https://arxiv.org/abs/1704.04854} {arXiv:1704.04854 [cond-mat.supr-con]} \BibitemShut {NoStop}%
\bibitem [{\citenamefont {Yao}\ \emph {et~al.}(2022)\citenamefont {Yao}, \citenamefont {Willa}, \citenamefont {Lacmann}, \citenamefont {Souliou}, \citenamefont {Frachet}, \citenamefont {Willa}, \citenamefont {Merz}, \citenamefont {Weber}, \citenamefont {Meingast}, \citenamefont {Heid}, \citenamefont {Haghighirad}, \citenamefont {Schmalian},\ and\ \citenamefont {Le~Tacon}}]{Yao2022}%
  \BibitemOpen
  \bibfield  {author} {\bibinfo {author} {\bibfnamefont {Y.}~\bibnamefont {Yao}}, \bibinfo {author} {\bibfnamefont {R.}~\bibnamefont {Willa}}, \bibinfo {author} {\bibfnamefont {T.}~\bibnamefont {Lacmann}}, \bibinfo {author} {\bibfnamefont {S.-M.}\ \bibnamefont {Souliou}}, \bibinfo {author} {\bibfnamefont {M.}~\bibnamefont {Frachet}}, \bibinfo {author} {\bibfnamefont {K.}~\bibnamefont {Willa}}, \bibinfo {author} {\bibfnamefont {M.}~\bibnamefont {Merz}}, \bibinfo {author} {\bibfnamefont {F.}~\bibnamefont {Weber}}, \bibinfo {author} {\bibfnamefont {C.}~\bibnamefont {Meingast}}, \bibinfo {author} {\bibfnamefont {R.}~\bibnamefont {Heid}}, \bibinfo {author} {\bibfnamefont {A.-A.}\ \bibnamefont {Haghighirad}}, \bibinfo {author} {\bibfnamefont {J.}~\bibnamefont {Schmalian}},\ and\ \bibinfo {author} {\bibfnamefont {M.}~\bibnamefont {Le~Tacon}},\ }\bibfield  {title} {\bibinfo {title} {An electronic nematic liquid in {BaNi$_2$As$_2$}},\ }\href {https://doi.org/10.1038/s41467-022-32112-7} {\bibfield  {journal} {\bibinfo
  {journal} {Nature Communications}\ }\textbf {\bibinfo {volume} {13}},\ \bibinfo {pages} {4535} (\bibinfo {year} {2022})}\BibitemShut {NoStop}%
\bibitem [{\citenamefont {Krisch}\ and\ \citenamefont {Sette}(2007)}]{Krisch2007}%
  \BibitemOpen
  \bibfield  {author} {\bibinfo {author} {\bibfnamefont {M.}~\bibnamefont {Krisch}}\ and\ \bibinfo {author} {\bibfnamefont {F.}~\bibnamefont {Sette}},\ }\bibinfo {title} {Inelastic x-ray scattering from phonons},\ in\ \href {https://doi.org/10.1007/978-3-540-34436-0_5} {\emph {\bibinfo {booktitle} {Light Scattering in Solid IX}}},\ \bibinfo {editor} {edited by\ \bibinfo {editor} {\bibfnamefont {M.}~\bibnamefont {Cardona}}\ and\ \bibinfo {editor} {\bibfnamefont {R.}~\bibnamefont {Merlin}}}\ (\bibinfo  {publisher} {Springer Berlin Heidelberg},\ \bibinfo {address} {Berlin, Heidelberg},\ \bibinfo {year} {2007})\ pp.\ \bibinfo {pages} {317--370}\BibitemShut {NoStop}%
\bibitem [{\citenamefont {Girard}\ \emph {et~al.}(2019)\citenamefont {Girard}, \citenamefont {Nguyen-Thanh}, \citenamefont {Souliou}, \citenamefont {Stekiel}, \citenamefont {Morgenroth}, \citenamefont {Paolasini}, \citenamefont {Minelli}, \citenamefont {Gambetti}, \citenamefont {Winkler},\ and\ \citenamefont {Bosak}}]{Girard2019}%
  \BibitemOpen
  \bibfield  {author} {\bibinfo {author} {\bibfnamefont {A.}~\bibnamefont {Girard}}, \bibinfo {author} {\bibfnamefont {T.}~\bibnamefont {Nguyen-Thanh}}, \bibinfo {author} {\bibfnamefont {S.~M.}\ \bibnamefont {Souliou}}, \bibinfo {author} {\bibfnamefont {M.}~\bibnamefont {Stekiel}}, \bibinfo {author} {\bibfnamefont {W.}~\bibnamefont {Morgenroth}}, \bibinfo {author} {\bibfnamefont {L.}~\bibnamefont {Paolasini}}, \bibinfo {author} {\bibfnamefont {A.}~\bibnamefont {Minelli}}, \bibinfo {author} {\bibfnamefont {D.}~\bibnamefont {Gambetti}}, \bibinfo {author} {\bibfnamefont {B.}~\bibnamefont {Winkler}},\ and\ \bibinfo {author} {\bibfnamefont {A.}~\bibnamefont {Bosak}},\ }\bibfield  {title} {\bibinfo {title} {{A new diffractometer for diffuse scattering studies on the ID28 beamline at the ESRF}},\ }\href {https://doi.org/10.1107/S1600577518016132} {\bibfield  {journal} {\bibinfo  {journal} {Journal of Synchrotron Radiation}\ }\textbf {\bibinfo {volume} {26}},\ \bibinfo {pages} {272} (\bibinfo {year}
  {2019})}\BibitemShut {NoStop}%
\bibitem [{Cry(2015)}]{CrysalysPro}%
  \BibitemOpen
  \href@noop {} {\bibinfo {title} {{Rigaku Oxford Diffraction Ltd, Yarnton, Oxfordshire, E 2015 CrysAlis PRO}}} (\bibinfo {year} {2015})\BibitemShut {NoStop}%
\bibitem [{SM()}]{SM}%
  \BibitemOpen
  \href@noop {} {}\bibinfo {note} {See Supplemental Material at [URL will be inserted by publisher] for information on the samples, additional reciprocal space maps, linecuts, fit parameters, IXS spectra around the superconducting transition and in the longitudinal direction, and a discussion of the indications of phonon softening around the position of a second I-CDW emerging under hydrostatic pressure.}\BibitemShut {Stop}%
\bibitem [{\citenamefont {Wehinger}\ \emph {et~al.}(2014)\citenamefont {Wehinger}, \citenamefont {Bosak}, \citenamefont {Piccolboni}, \citenamefont {Refson}, \citenamefont {Chernyshov}, \citenamefont {Ivanov}, \citenamefont {Rumiantsev},\ and\ \citenamefont {Krisch}}]{Wehinger2014}%
  \BibitemOpen
  \bibfield  {author} {\bibinfo {author} {\bibfnamefont {B.}~\bibnamefont {Wehinger}}, \bibinfo {author} {\bibfnamefont {A.}~\bibnamefont {Bosak}}, \bibinfo {author} {\bibfnamefont {G.}~\bibnamefont {Piccolboni}}, \bibinfo {author} {\bibfnamefont {K.}~\bibnamefont {Refson}}, \bibinfo {author} {\bibfnamefont {D.}~\bibnamefont {Chernyshov}}, \bibinfo {author} {\bibfnamefont {A.}~\bibnamefont {Ivanov}}, \bibinfo {author} {\bibfnamefont {A.}~\bibnamefont {Rumiantsev}},\ and\ \bibinfo {author} {\bibfnamefont {M.}~\bibnamefont {Krisch}},\ }\bibfield  {title} {\bibinfo {title} {Diffuse scattering in metallic tin polymorphs},\ }\href {https://doi.org/10.1088/0953-8984/26/11/115401} {\bibfield  {journal} {\bibinfo  {journal} {Journal of Physics: Condensed Matter}\ }\textbf {\bibinfo {volume} {26}},\ \bibinfo {pages} {115401} (\bibinfo {year} {2014})}\BibitemShut {NoStop}%
\bibitem [{\citenamefont {Mirone}\ and\ \citenamefont {Wehinger}()}]{soft:ab2tds}%
  \BibitemOpen
  \bibfield  {author} {\bibinfo {author} {\bibfnamefont {A.}~\bibnamefont {Mirone}}\ and\ \bibinfo {author} {\bibfnamefont {B.}~\bibnamefont {Wehinger}},\ }\href@noop {} {\bibinfo {title} {ab2tds, version 1.1, {European Synchrotron Radiation Facility}}}\BibitemShut {NoStop}%
\bibitem [{\citenamefont {Eagen}\ \emph {et~al.}(1975)\citenamefont {Eagen}, \citenamefont {Werner},\ and\ \citenamefont {Saillant}}]{Eagen1975}%
  \BibitemOpen
  \bibfield  {author} {\bibinfo {author} {\bibfnamefont {C.~F.}\ \bibnamefont {Eagen}}, \bibinfo {author} {\bibfnamefont {S.~A.}\ \bibnamefont {Werner}},\ and\ \bibinfo {author} {\bibfnamefont {R.~B.}\ \bibnamefont {Saillant}},\ }\bibfield  {title} {\bibinfo {title} {Amplitude and nature of the charge-density-wave displacements in {${\mathrm{K}}_{2}$Pt${(\mathrm{CN})}_{4}$${\mathrm{Br}}_{0.3}$.3.2${\mathrm{D}}_{2}$O} ({KCP}) at low temperatures},\ }\href {https://doi.org/10.1103/PhysRevB.12.2036} {\bibfield  {journal} {\bibinfo  {journal} {Phys. Rev. B}\ }\textbf {\bibinfo {volume} {12}},\ \bibinfo {pages} {2036} (\bibinfo {year} {1975})}\BibitemShut {NoStop}%
\bibitem [{\citenamefont {Song}\ \emph {et~al.}(2003)\citenamefont {Song}, \citenamefont {Park}, \citenamefont {Koo}, \citenamefont {Lee}, \citenamefont {Rhee}, \citenamefont {Bud'ko}, \citenamefont {Canfield}, \citenamefont {Harmon},\ and\ \citenamefont {Goldman}}]{Song2003}%
  \BibitemOpen
  \bibfield  {author} {\bibinfo {author} {\bibfnamefont {C.}~\bibnamefont {Song}}, \bibinfo {author} {\bibfnamefont {J.}~\bibnamefont {Park}}, \bibinfo {author} {\bibfnamefont {J.}~\bibnamefont {Koo}}, \bibinfo {author} {\bibfnamefont {K.-B.}\ \bibnamefont {Lee}}, \bibinfo {author} {\bibfnamefont {J.~Y.}\ \bibnamefont {Rhee}}, \bibinfo {author} {\bibfnamefont {S.~L.}\ \bibnamefont {Bud'ko}}, \bibinfo {author} {\bibfnamefont {P.~C.}\ \bibnamefont {Canfield}}, \bibinfo {author} {\bibfnamefont {B.~N.}\ \bibnamefont {Harmon}},\ and\ \bibinfo {author} {\bibfnamefont {A.~I.}\ \bibnamefont {Goldman}},\ }\bibfield  {title} {\bibinfo {title} {Charge-density-wave orderings in {${\mathrm{LaAgSb}}_{2}:$} an x-ray scattering study},\ }\href {https://doi.org/10.1103/PhysRevB.68.035113} {\bibfield  {journal} {\bibinfo  {journal} {Phys. Rev. B}\ }\textbf {\bibinfo {volume} {68}},\ \bibinfo {pages} {035113} (\bibinfo {year} {2003})}\BibitemShut {NoStop}%
\bibitem [{\citenamefont {Bosak}\ \emph {et~al.}(2021)\citenamefont {Bosak}, \citenamefont {Souliou}, \citenamefont {Faugeras}, \citenamefont {Heid}, \citenamefont {Molas}, \citenamefont {Chen}, \citenamefont {Wang}, \citenamefont {Potemski},\ and\ \citenamefont {Le~Tacon}}]{Bosak2021}%
  \BibitemOpen
  \bibfield  {author} {\bibinfo {author} {\bibfnamefont {A.}~\bibnamefont {Bosak}}, \bibinfo {author} {\bibfnamefont {S.-M.}\ \bibnamefont {Souliou}}, \bibinfo {author} {\bibfnamefont {C.}~\bibnamefont {Faugeras}}, \bibinfo {author} {\bibfnamefont {R.}~\bibnamefont {Heid}}, \bibinfo {author} {\bibfnamefont {M.~R.}\ \bibnamefont {Molas}}, \bibinfo {author} {\bibfnamefont {R.-Y.}\ \bibnamefont {Chen}}, \bibinfo {author} {\bibfnamefont {N.-L.}\ \bibnamefont {Wang}}, \bibinfo {author} {\bibfnamefont {M.}~\bibnamefont {Potemski}},\ and\ \bibinfo {author} {\bibfnamefont {M.}~\bibnamefont {Le~Tacon}},\ }\bibfield  {title} {\bibinfo {title} {Evidence for nesting-driven charge density wave instabilities in the quasi-two-dimensional material {${\mathrm{LaAgSb}}_{2}$}},\ }\href {https://doi.org/10.1103/PhysRevResearch.3.033020} {\bibfield  {journal} {\bibinfo  {journal} {Physical Review Research}\ }\textbf {\bibinfo {volume} {3}},\ \bibinfo {pages} {033020} (\bibinfo {year} {2021})}\BibitemShut {NoStop}%
\bibitem [{\citenamefont {Pokharel}\ \emph {et~al.}(2022)\citenamefont {Pokharel}, \citenamefont {Grigorev}, \citenamefont {Mejas}, \citenamefont {Dong}, \citenamefont {Haghighirad}, \citenamefont {Heid}, \citenamefont {Yao}, \citenamefont {Merz}, \citenamefont {Le~Tacon},\ and\ \citenamefont {Demsar}}]{Pokharel2022}%
  \BibitemOpen
  \bibfield  {author} {\bibinfo {author} {\bibfnamefont {A.~R.}\ \bibnamefont {Pokharel}}, \bibinfo {author} {\bibfnamefont {V.}~\bibnamefont {Grigorev}}, \bibinfo {author} {\bibfnamefont {A.}~\bibnamefont {Mejas}}, \bibinfo {author} {\bibfnamefont {T.}~\bibnamefont {Dong}}, \bibinfo {author} {\bibfnamefont {A.~A.}\ \bibnamefont {Haghighirad}}, \bibinfo {author} {\bibfnamefont {R.}~\bibnamefont {Heid}}, \bibinfo {author} {\bibfnamefont {Y.}~\bibnamefont {Yao}}, \bibinfo {author} {\bibfnamefont {M.}~\bibnamefont {Merz}}, \bibinfo {author} {\bibfnamefont {M.}~\bibnamefont {Le~Tacon}},\ and\ \bibinfo {author} {\bibfnamefont {J.}~\bibnamefont {Demsar}},\ }\bibfield  {title} {\bibinfo {title} {Dynamics of collective modes in an unconventional charge density wave system {BaNi\textsubscript{2}As\textsubscript{2}}},\ }\href {https://doi.org/10.1038/s42005-022-00919-x} {\bibfield  {journal} {\bibinfo  {journal} {Communications Physics}\ }\textbf {\bibinfo {volume} {5}},\ \bibinfo {pages} {141} (\bibinfo {year}
  {2022})}\BibitemShut {NoStop}%
\bibitem [{\citenamefont {Chen}\ \emph {et~al.}(2024)\citenamefont {Chen}, \citenamefont {Giles-Donovan}, \citenamefont {Guo}, \citenamefont {Chen}, \citenamefont {Fukui}, \citenamefont {Manjo}, \citenamefont {Ishikawa}, \citenamefont {Baron}, \citenamefont {Song},\ and\ \citenamefont {Birgeneau}}]{chen2024}%
  \BibitemOpen
  \bibfield  {author} {\bibinfo {author} {\bibfnamefont {Y.}~\bibnamefont {Chen}}, \bibinfo {author} {\bibfnamefont {N.}~\bibnamefont {Giles-Donovan}}, \bibinfo {author} {\bibfnamefont {J.}~\bibnamefont {Guo}}, \bibinfo {author} {\bibfnamefont {R.}~\bibnamefont {Chen}}, \bibinfo {author} {\bibfnamefont {H.}~\bibnamefont {Fukui}}, \bibinfo {author} {\bibfnamefont {T.}~\bibnamefont {Manjo}}, \bibinfo {author} {\bibfnamefont {D.}~\bibnamefont {Ishikawa}}, \bibinfo {author} {\bibfnamefont {A.~Q.~R.}\ \bibnamefont {Baron}}, \bibinfo {author} {\bibfnamefont {Y.}~\bibnamefont {Song}},\ and\ \bibinfo {author} {\bibfnamefont {R.~J.}\ \bibnamefont {Birgeneau}},\ }\bibfield  {title} {\bibinfo {title} {Charge density fluctuations with enhanced superconductivity at the proposed nematic quantum critical point},\ }\Eprint {https://arxiv.org/abs/2410.03956} {arXiv:2410.03956 [cond-mat.supr-con]}  (\bibinfo {year} {2024})\BibitemShut {NoStop}%
\bibitem [{\citenamefont {Ghiringhelli}\ \emph {et~al.}(2012)\citenamefont {Ghiringhelli}, \citenamefont {Tacon}, \citenamefont {Minola}, \citenamefont {Blanco-Canosa}, \citenamefont {Mazzoli}, \citenamefont {Brookes}, \citenamefont {Luca}, \citenamefont {Frano}, \citenamefont {Hawthorn}, \citenamefont {He}, \citenamefont {Loew}, \citenamefont {Sala}, \citenamefont {Peets}, \citenamefont {Salluzzo}, \citenamefont {Schierle}, \citenamefont {Sutarto}, \citenamefont {Sawatzky}, \citenamefont {Weschke}, \citenamefont {Keimer},\ and\ \citenamefont {Braicovich}}]{Ghiringhelli2012}%
  \BibitemOpen
  \bibfield  {author} {\bibinfo {author} {\bibfnamefont {G.}~\bibnamefont {Ghiringhelli}}, \bibinfo {author} {\bibfnamefont {M.~L.}\ \bibnamefont {Tacon}}, \bibinfo {author} {\bibfnamefont {M.}~\bibnamefont {Minola}}, \bibinfo {author} {\bibfnamefont {S.}~\bibnamefont {Blanco-Canosa}}, \bibinfo {author} {\bibfnamefont {C.}~\bibnamefont {Mazzoli}}, \bibinfo {author} {\bibfnamefont {N.~B.}\ \bibnamefont {Brookes}}, \bibinfo {author} {\bibfnamefont {G.~M.~D.}\ \bibnamefont {Luca}}, \bibinfo {author} {\bibfnamefont {A.}~\bibnamefont {Frano}}, \bibinfo {author} {\bibfnamefont {D.~G.}\ \bibnamefont {Hawthorn}}, \bibinfo {author} {\bibfnamefont {F.}~\bibnamefont {He}}, \bibinfo {author} {\bibfnamefont {T.}~\bibnamefont {Loew}}, \bibinfo {author} {\bibfnamefont {M.~M.}\ \bibnamefont {Sala}}, \bibinfo {author} {\bibfnamefont {D.~C.}\ \bibnamefont {Peets}}, \bibinfo {author} {\bibfnamefont {M.}~\bibnamefont {Salluzzo}}, \bibinfo {author} {\bibfnamefont {E.}~\bibnamefont {Schierle}}, \bibinfo {author} {\bibfnamefont
  {R.}~\bibnamefont {Sutarto}}, \bibinfo {author} {\bibfnamefont {G.~A.}\ \bibnamefont {Sawatzky}}, \bibinfo {author} {\bibfnamefont {E.}~\bibnamefont {Weschke}}, \bibinfo {author} {\bibfnamefont {B.}~\bibnamefont {Keimer}},\ and\ \bibinfo {author} {\bibfnamefont {L.}~\bibnamefont {Braicovich}},\ }\bibfield  {title} {\bibinfo {title} {Long-range incommensurate charge fluctuations in {(Y,Nd)Ba\textsubscript{2}Cu\textsubscript{3}O\textsubscript{6+x}}},\ }\href {https://doi.org/10.1126/science.1223532} {\bibfield  {journal} {\bibinfo  {journal} {Science}\ }\textbf {\bibinfo {volume} {337}},\ \bibinfo {pages} {821} (\bibinfo {year} {2012})}\BibitemShut {NoStop}%
\bibitem [{\citenamefont {Chang}\ \emph {et~al.}(2012)\citenamefont {Chang}, \citenamefont {Blackburn}, \citenamefont {Holmes}, \citenamefont {Christensen}, \citenamefont {Larsen}, \citenamefont {Mesot}, \citenamefont {Liang}, \citenamefont {Bonn}, \citenamefont {Hardy}, \citenamefont {Watenphul}, \citenamefont {Zimmermann}, \citenamefont {Forgan},\ and\ \citenamefont {Hayden}}]{Chang2012}%
  \BibitemOpen
  \bibfield  {author} {\bibinfo {author} {\bibfnamefont {J.}~\bibnamefont {Chang}}, \bibinfo {author} {\bibfnamefont {E.}~\bibnamefont {Blackburn}}, \bibinfo {author} {\bibfnamefont {A.~T.}\ \bibnamefont {Holmes}}, \bibinfo {author} {\bibfnamefont {N.~B.}\ \bibnamefont {Christensen}}, \bibinfo {author} {\bibfnamefont {J.}~\bibnamefont {Larsen}}, \bibinfo {author} {\bibfnamefont {J.}~\bibnamefont {Mesot}}, \bibinfo {author} {\bibfnamefont {R.}~\bibnamefont {Liang}}, \bibinfo {author} {\bibfnamefont {D.~A.}\ \bibnamefont {Bonn}}, \bibinfo {author} {\bibfnamefont {W.~N.}\ \bibnamefont {Hardy}}, \bibinfo {author} {\bibfnamefont {A.}~\bibnamefont {Watenphul}}, \bibinfo {author} {\bibfnamefont {M.~v.}\ \bibnamefont {Zimmermann}}, \bibinfo {author} {\bibfnamefont {E.~M.}\ \bibnamefont {Forgan}},\ and\ \bibinfo {author} {\bibfnamefont {S.~M.}\ \bibnamefont {Hayden}},\ }\bibfield  {title} {\bibinfo {title} {Direct observation of competition between superconductivity and charge density wave order in
  {YBa\textsubscript{2}Cu\textsubscript{3}O\textsubscript{6.67}}},\ }\href {https://doi.org/10.1038/nphys2456} {\bibfield  {journal} {\bibinfo  {journal} {Nature Physics}\ }\textbf {\bibinfo {volume} {8}},\ \bibinfo {pages} {871} (\bibinfo {year} {2012})}\BibitemShut {NoStop}%
\bibitem [{\citenamefont {Vinograd}\ \emph {et~al.}(2024)\citenamefont {Vinograd}, \citenamefont {Souliou}, \citenamefont {Haghighirad}, \citenamefont {Lacmann}, \citenamefont {Caplan}, \citenamefont {Frachet}, \citenamefont {Merz}, \citenamefont {Garbarino}, \citenamefont {Liu}, \citenamefont {Nakata}, \citenamefont {Ishida}, \citenamefont {Noad}, \citenamefont {Minola}, \citenamefont {Keimer}, \citenamefont {Orgad}, \citenamefont {Hicks},\ and\ \citenamefont {Le~Tacon}}]{Vinograd2024}%
  \BibitemOpen
  \bibfield  {author} {\bibinfo {author} {\bibfnamefont {I.}~\bibnamefont {Vinograd}}, \bibinfo {author} {\bibfnamefont {S.~M.}\ \bibnamefont {Souliou}}, \bibinfo {author} {\bibfnamefont {A.-A.}\ \bibnamefont {Haghighirad}}, \bibinfo {author} {\bibfnamefont {T.}~\bibnamefont {Lacmann}}, \bibinfo {author} {\bibfnamefont {Y.}~\bibnamefont {Caplan}}, \bibinfo {author} {\bibfnamefont {M.}~\bibnamefont {Frachet}}, \bibinfo {author} {\bibfnamefont {M.}~\bibnamefont {Merz}}, \bibinfo {author} {\bibfnamefont {G.}~\bibnamefont {Garbarino}}, \bibinfo {author} {\bibfnamefont {Y.}~\bibnamefont {Liu}}, \bibinfo {author} {\bibfnamefont {S.}~\bibnamefont {Nakata}}, \bibinfo {author} {\bibfnamefont {K.}~\bibnamefont {Ishida}}, \bibinfo {author} {\bibfnamefont {H.~M.~L.}\ \bibnamefont {Noad}}, \bibinfo {author} {\bibfnamefont {M.}~\bibnamefont {Minola}}, \bibinfo {author} {\bibfnamefont {B.}~\bibnamefont {Keimer}}, \bibinfo {author} {\bibfnamefont {D.}~\bibnamefont {Orgad}}, \bibinfo {author} {\bibfnamefont {C.~W.}\
  \bibnamefont {Hicks}},\ and\ \bibinfo {author} {\bibfnamefont {M.}~\bibnamefont {Le~Tacon}},\ }\bibfield  {title} {\bibinfo {title} {Using strain to uncover the interplay between two- and three-dimensional charge density waves in high-temperature superconducting {YBa\textsubscript{2}Cu\textsubscript{3}O\textsubscript{y}}},\ }\href {https://doi.org/10.1038/s41467-024-47540-w} {\bibfield  {journal} {\bibinfo  {journal} {Nature Communications}\ }\textbf {\bibinfo {volume} {15}},\ \bibinfo {pages} {3277} (\bibinfo {year} {2024})}\BibitemShut {NoStop}%
\bibitem [{\citenamefont {Leroux}\ \emph {et~al.}(2020)\citenamefont {Leroux}, \citenamefont {Mishra}, \citenamefont {Opagiste}, \citenamefont {Rodi\`ere}, \citenamefont {Kayani}, \citenamefont {Kwok},\ and\ \citenamefont {Welp}}]{Leroux2020}%
  \BibitemOpen
  \bibfield  {author} {\bibinfo {author} {\bibfnamefont {M.}~\bibnamefont {Leroux}}, \bibinfo {author} {\bibfnamefont {V.}~\bibnamefont {Mishra}}, \bibinfo {author} {\bibfnamefont {C.}~\bibnamefont {Opagiste}}, \bibinfo {author} {\bibfnamefont {P.}~\bibnamefont {Rodi\`ere}}, \bibinfo {author} {\bibfnamefont {A.}~\bibnamefont {Kayani}}, \bibinfo {author} {\bibfnamefont {W.-K.}\ \bibnamefont {Kwok}},\ and\ \bibinfo {author} {\bibfnamefont {U.}~\bibnamefont {Welp}},\ }\bibfield  {title} {\bibinfo {title} {Charge density wave and superconductivity competition in {${\mathrm{Lu}}_{5}{\mathrm{Ir}}_{4}{\mathrm{Si}}_{10}$}: A proton irradiation study},\ }\href {https://doi.org/10.1103/PhysRevB.102.094519} {\bibfield  {journal} {\bibinfo  {journal} {Phys. Rev. B}\ }\textbf {\bibinfo {volume} {102}},\ \bibinfo {pages} {094519} (\bibinfo {year} {2020})}\BibitemShut {NoStop}%
\bibitem [{\citenamefont {Stier}\ \emph {et~al.}(2024)\citenamefont {Stier}, \citenamefont {Haghighirad}, \citenamefont {Garbarino}, \citenamefont {Mishra}, \citenamefont {Stilkerich}, \citenamefont {Chen}, \citenamefont {Shekhar}, \citenamefont {Lacmann}, \citenamefont {Felser}, \citenamefont {Ritschel}, \citenamefont {Geck},\ and\ \citenamefont {Le~Tacon}}]{Stier2024}%
  \BibitemOpen
  \bibfield  {author} {\bibinfo {author} {\bibfnamefont {F.}~\bibnamefont {Stier}}, \bibinfo {author} {\bibfnamefont {A.-A.}\ \bibnamefont {Haghighirad}}, \bibinfo {author} {\bibfnamefont {G.}~\bibnamefont {Garbarino}}, \bibinfo {author} {\bibfnamefont {S.}~\bibnamefont {Mishra}}, \bibinfo {author} {\bibfnamefont {N.}~\bibnamefont {Stilkerich}}, \bibinfo {author} {\bibfnamefont {D.}~\bibnamefont {Chen}}, \bibinfo {author} {\bibfnamefont {C.}~\bibnamefont {Shekhar}}, \bibinfo {author} {\bibfnamefont {T.}~\bibnamefont {Lacmann}}, \bibinfo {author} {\bibfnamefont {C.}~\bibnamefont {Felser}}, \bibinfo {author} {\bibfnamefont {T.}~\bibnamefont {Ritschel}}, \bibinfo {author} {\bibfnamefont {J.}~\bibnamefont {Geck}},\ and\ \bibinfo {author} {\bibfnamefont {M.}~\bibnamefont {Le~Tacon}},\ }\bibfield  {title} {\bibinfo {title} {Pressure-dependent electronic superlattice in the kagome superconductor {${\mathrm{CsV}}_{3}{\mathrm{Sb}}_{5}$}},\ }\href {https://doi.org/10.1103/PhysRevLett.133.236503} {\bibfield  {journal}
  {\bibinfo  {journal} {Phys. Rev. Lett.}\ }\textbf {\bibinfo {volume} {133}},\ \bibinfo {pages} {236503} (\bibinfo {year} {2024})}\BibitemShut {NoStop}%
\bibitem [{\citenamefont {Lacmann}\ \emph {et~al.}(2025)\citenamefont {Lacmann}, \citenamefont {Souliou}, \citenamefont {Henssler}, \citenamefont {Frachet}, \citenamefont {McGuinness}, \citenamefont {Chaney},\ and\ \citenamefont {Le~Tacon}}]{Data:KIT}%
  \BibitemOpen
  \bibfield  {author} {\bibinfo {author} {\bibfnamefont {T.}~\bibnamefont {Lacmann}}, \bibinfo {author} {\bibfnamefont {S.-M.}\ \bibnamefont {Souliou}}, \bibinfo {author} {\bibfnamefont {F.}~\bibnamefont {Henssler}}, \bibinfo {author} {\bibfnamefont {M.}~\bibnamefont {Frachet}}, \bibinfo {author} {\bibfnamefont {P.~H.}\ \bibnamefont {McGuinness}}, \bibinfo {author} {\bibfnamefont {D.~A.}\ \bibnamefont {Chaney}},\ and\ \bibinfo {author} {\bibfnamefont {M.}~\bibnamefont {Le~Tacon}},\ }\bibfield  {title} {\bibinfo {title} {Charge density waves and soft phonon evolution in the superconductor {BaNi\textsubscript{2}(As\textsubscript{1-\textit{x}}P\textsubscript{\textit{x}})\textsubscript{2}} [data set]},\ }\href {https://doi.org/10.35097/485k5eg675xf4c6d} {10.35097/485k5eg675xf4c6d} (\bibinfo {year} {2025}),\ \bibinfo {note} {{Karlsruhe Institute of Technology}}\BibitemShut {NoStop}%
\end{thebibliography}%


\providecommand{\noopsort}[1]{}\providecommand{\singleletter}[1]{#1}%
\begin{thebibliography}{3}%
\makeatletter
\providecommand \@ifxundefined [1]{%
 \@ifx{#1\undefined}
}%
\providecommand \@ifnum [1]{%
 \ifnum #1\expandafter \@firstoftwo
 \else \expandafter \@secondoftwo
 \fi
}%
\providecommand \@ifx [1]{%
 \ifx #1\expandafter \@firstoftwo
 \else \expandafter \@secondoftwo
 \fi
}%
\providecommand \natexlab [1]{#1}%
\providecommand \enquote  [1]{``#1''}%
\providecommand \bibnamefont  [1]{#1}%
\providecommand \bibfnamefont [1]{#1}%
\providecommand \citenamefont [1]{#1}%
\providecommand \href@noop [0]{\@secondoftwo}%
\providecommand \href [0]{\begingroup \@sanitize@url \@href}%
\providecommand \@href[1]{\@@startlink{#1}\@@href}%
\providecommand \@@href[1]{\endgroup#1\@@endlink}%
\providecommand \@sanitize@url [0]{\catcode `\\12\catcode `\$12\catcode `\&12\catcode `\#12\catcode `\^12\catcode `\_12\catcode `\%12\relax}%
\providecommand \@@startlink[1]{}%
\providecommand \@@endlink[0]{}%
\providecommand \url  [0]{\begingroup\@sanitize@url \@url }%
\providecommand \@url [1]{\endgroup\@href {#1}{\urlprefix }}%
\providecommand \urlprefix  [0]{URL }%
\providecommand \Eprint [0]{\href }%
\providecommand \doibase [0]{https://doi.org/}%
\providecommand \selectlanguage [0]{\@gobble}%
\providecommand \bibinfo  [0]{\@secondoftwo}%
\providecommand \bibfield  [0]{\@secondoftwo}%
\providecommand \translation [1]{[#1]}%
\providecommand \BibitemOpen [0]{}%
\providecommand \bibitemStop [0]{}%
\providecommand \bibitemNoStop [0]{.\EOS\space}%
\providecommand \EOS [0]{\spacefactor3000\relax}%
\providecommand \BibitemShut  [1]{\csname bibitem#1\endcsname}%
\let\auto@bib@innerbib\@empty
\bibitem [{\citenamefont {Meingast}\ \emph {et~al.}(2022)\citenamefont {Meingast}, \citenamefont {Shukla}, \citenamefont {Wang}, \citenamefont {Heid}, \citenamefont {Hardy}, \citenamefont {Frachet}, \citenamefont {Willa}, \citenamefont {Lacmann}, \citenamefont {Le~Tacon}, \citenamefont {Merz}, \citenamefont {Haghighirad},\ and\ \citenamefont {Wolf}}]{Meingast2022}%
  \BibitemOpen
  \bibfield  {author} {\bibinfo {author} {\bibfnamefont {C.}~\bibnamefont {Meingast}}, \bibinfo {author} {\bibfnamefont {A.}~\bibnamefont {Shukla}}, \bibinfo {author} {\bibfnamefont {L.}~\bibnamefont {Wang}}, \bibinfo {author} {\bibfnamefont {R.}~\bibnamefont {Heid}}, \bibinfo {author} {\bibfnamefont {F.}~\bibnamefont {Hardy}}, \bibinfo {author} {\bibfnamefont {M.}~\bibnamefont {Frachet}}, \bibinfo {author} {\bibfnamefont {K.}~\bibnamefont {Willa}}, \bibinfo {author} {\bibfnamefont {T.}~\bibnamefont {Lacmann}}, \bibinfo {author} {\bibfnamefont {M.}~\bibnamefont {Le~Tacon}}, \bibinfo {author} {\bibfnamefont {M.}~\bibnamefont {Merz}}, \bibinfo {author} {\bibfnamefont {A.-A.}\ \bibnamefont {Haghighirad}},\ and\ \bibinfo {author} {\bibfnamefont {T.}~\bibnamefont {Wolf}},\ }\bibfield  {title} {\bibinfo {title} {Charge density wave transitions, soft phonon, and possible electronic nematicity in {${\mathrm{BaNi}}_{2}({\mathrm{As}}_{1\ensuremath{-}x}{\mathrm{P}}_{x}{)}_{2}$}},\ }\href
  {https://doi.org/10.1103/PhysRevB.106.144507} {\bibfield  {journal} {\bibinfo  {journal} {Phys. Rev. B}\ }\textbf {\bibinfo {volume} {106}},\ \bibinfo {pages} {144507} (\bibinfo {year} {2022})}\BibitemShut {NoStop}%
\bibitem [{\citenamefont {Souliou}\ \emph {et~al.}(2022)\citenamefont {Souliou}, \citenamefont {Lacmann}, \citenamefont {Heid}, \citenamefont {Meingast}, \citenamefont {Frachet}, \citenamefont {Paolasini}, \citenamefont {Haghighirad}, \citenamefont {Merz}, \citenamefont {Bosak},\ and\ \citenamefont {Le~Tacon}}]{Souliou2022}%
  \BibitemOpen
  \bibfield  {author} {\bibinfo {author} {\bibfnamefont {S.~M.}\ \bibnamefont {Souliou}}, \bibinfo {author} {\bibfnamefont {T.}~\bibnamefont {Lacmann}}, \bibinfo {author} {\bibfnamefont {R.}~\bibnamefont {Heid}}, \bibinfo {author} {\bibfnamefont {C.}~\bibnamefont {Meingast}}, \bibinfo {author} {\bibfnamefont {M.}~\bibnamefont {Frachet}}, \bibinfo {author} {\bibfnamefont {L.}~\bibnamefont {Paolasini}}, \bibinfo {author} {\bibfnamefont {A.-A.}\ \bibnamefont {Haghighirad}}, \bibinfo {author} {\bibfnamefont {M.}~\bibnamefont {Merz}}, \bibinfo {author} {\bibfnamefont {A.}~\bibnamefont {Bosak}},\ and\ \bibinfo {author} {\bibfnamefont {M.}~\bibnamefont {Le~Tacon}},\ }\bibfield  {title} {\bibinfo {title} {Soft-phonon and charge-density-wave formation in nematic {${\mathrm{BaNi}}_{2}{\mathrm{As}}_{2}$}},\ }\href {https://doi.org/10.1103/PhysRevLett.129.247602} {\bibfield  {journal} {\bibinfo  {journal} {Phys. Rev. Lett.}\ }\textbf {\bibinfo {volume} {129}},\ \bibinfo {pages} {247602} (\bibinfo {year}
  {2022})}\BibitemShut {NoStop}%
\bibitem [{\citenamefont {Lacmann}\ \emph {et~al.}(2023)\citenamefont {Lacmann}, \citenamefont {Haghighirad}, \citenamefont {Souliou}, \citenamefont {Merz}, \citenamefont {Garbarino}, \citenamefont {Glazyrin}, \citenamefont {Heid},\ and\ \citenamefont {Le~Tacon}}]{Lacmann2023}%
  \BibitemOpen
  \bibfield  {author} {\bibinfo {author} {\bibfnamefont {T.}~\bibnamefont {Lacmann}}, \bibinfo {author} {\bibfnamefont {A.-A.}\ \bibnamefont {Haghighirad}}, \bibinfo {author} {\bibfnamefont {S.-M.}\ \bibnamefont {Souliou}}, \bibinfo {author} {\bibfnamefont {M.}~\bibnamefont {Merz}}, \bibinfo {author} {\bibfnamefont {G.}~\bibnamefont {Garbarino}}, \bibinfo {author} {\bibfnamefont {K.}~\bibnamefont {Glazyrin}}, \bibinfo {author} {\bibfnamefont {R.}~\bibnamefont {Heid}},\ and\ \bibinfo {author} {\bibfnamefont {M.}~\bibnamefont {Le~Tacon}},\ }\bibfield  {title} {\bibinfo {title} {High-pressure phase diagram of {${\mathrm{BaNi}}_{2}{\mathrm{As}}_{2}$}: Unconventional charge density waves and structural phase transitions},\ }\href {https://doi.org/10.1103/PhysRevB.108.224115} {\bibfield  {journal} {\bibinfo  {journal} {Phys. Rev. B}\ }\textbf {\bibinfo {volume} {108}},\ \bibinfo {pages} {224115} (\bibinfo {year} {2023})}\BibitemShut {NoStop}%
\end{thebibliography}%

\end{document}


\newcommand*{\titlename}{Charge density waves and soft phonon evolution in the superconductor BaNi\textsubscript{2}(As\textsubscript{1-\textit{x}}P\textsubscript{\textit{x}})\textsubscript{2}}
\title{Supplemental Material: \titlename}
\author{Tom Lacmann}
\affiliation{Institute for Quantum Materials and Technologies, Karlsruhe Institute of Technology, 76021 Karlsruhe, Germany}
\author{Sofia-Michaela Souliou}
\affiliation{Institute for Quantum Materials and Technologies, Karlsruhe Institute of Technology, 76021 Karlsruhe, Germany}
\author{Fabian Henssler}
\affiliation{Institute for Quantum Materials and Technologies, Karlsruhe Institute of Technology, 76021 Karlsruhe, Germany}
\author{Mehdi Frachet}
\affiliation{Institute for Quantum Materials and Technologies, Karlsruhe Institute of Technology, 76021 Karlsruhe, Germany}
\author{Philippa Helen McGuinness}
\affiliation{Institute for Quantum Materials and Technologies, Karlsruhe Institute of Technology, 76021 Karlsruhe, Germany}
\author{Michael Merz}
\affiliation{Institute for Quantum Materials and Technologies, Karlsruhe Institute of Technology, 76021 Karlsruhe, Germany}
\affiliation{Karlsruhe Nano Micro Facility (KNMFi), Karlsruhe Institute of Technology, Kaiserstr. 12, 76131 Karlsruhe, Germany}
\author{Björn Wehinger}
\affiliation{ESRF, The European Synchrotron, 71, avenue des Martyrs, CS 40220 F-38043 Grenoble Cedex 9}
\author{Daniel A. Chaney}
\affiliation{ESRF, The European Synchrotron, 71, avenue des Martyrs, CS 40220 F-38043 Grenoble Cedex 9}
\author{Amir-Abbas Haghighirad}
\affiliation{Institute for Quantum Materials and Technologies, Karlsruhe Institute of Technology, 76021 Karlsruhe, Germany}
\author{Rolf Heid}
\affiliation{Institute for Quantum Materials and Technologies, Karlsruhe Institute of Technology, 76021 Karlsruhe, Germany}
\author{Matthieu Le Tacon}
\email{Matthieu.LeTacon@kit.edu}
\affiliation{Institute for Quantum Materials and Technologies, Karlsruhe Institute of Technology, 76021 Karlsruhe, Germany}
\date{\today}
\maketitle

\section{Samples}

\newcolumntype{Y}{>{\centering\arraybackslash}X}
\begin{table*}
    \centering
    \caption{Summary of the samples, phosphorus content, sample mounting and used cryostat for the thermal diffuse X-ray scattering and inelastic X-ray scattering experiments.}
    \label{tab:samples}
    \begin{tabularx}{\textwidth}{X|YYY|YYYYY}
         &  \multicolumn{3}{c|}{IXS}  & \multicolumn{5}{c}{TDS} \\
          &  sample 1 & sample 2 & sample 3 & sample 4 & sample 5 & sample 6 & sample 7 & sample 8\\
         \hline
         Nominal \(x\) & 0.1 & 0.1 & 0.035 & 0.035 & 0.05 & 0.07 & 0.1 & 0.1\\
         \hline
         Real \(x\) & - & 0.107(6) & 0.0366(17) & 0.036(8) & 0.063(4) & 0.066(4) & 0.090(3) & 0.12(1)\\
         \hline
         Mounting & Araldite and silverpaint on copper rod & Silver epoxy (H20E) on copper rod & Araldite on glass needle & Araldite on glass needle & Araldite on glass needle & Araldite on glass needle & Araldite on glass needle & Silver epoxy (H20E) on copper rod\\
         \hline
         Cryostat & Joule-Thompson Displex & Joule-Thompson Displex & Cryostream & Cryostream & Cryostream &  Cryostream & Cryostream & Displex\\
    \end{tabularx}
\end{table*}

In total, five different samples for diffuse X-ray scattering experiments and an additional three different samples for inelastic X-ray scattering experiments with four different substitution levels were investigated. Samples with similar composition were taken from the same original piece if possible, or at least from the same batch. The samples are summarized in \cref{tab:samples}. The stated phosphorus contents were measured for all samples, except for the first IXS sample, with energy dispersive x-ray spectroscopy.  More details on the growth and composition determination are given in the methods section. No significant discrepancy in the IXS and TDS measurements was observed for the different samples. Details of the crystal structure of from these samples - or ones from the same batches - were previously obtained using x-ray diffraction and reported in ref.~\cite{Meingast2022}.

\FloatBarrier
\section{I-CDW Diffuse Scattering}

\subsection{Reciprocal space maps}

\begin{figure*}
\includegraphics{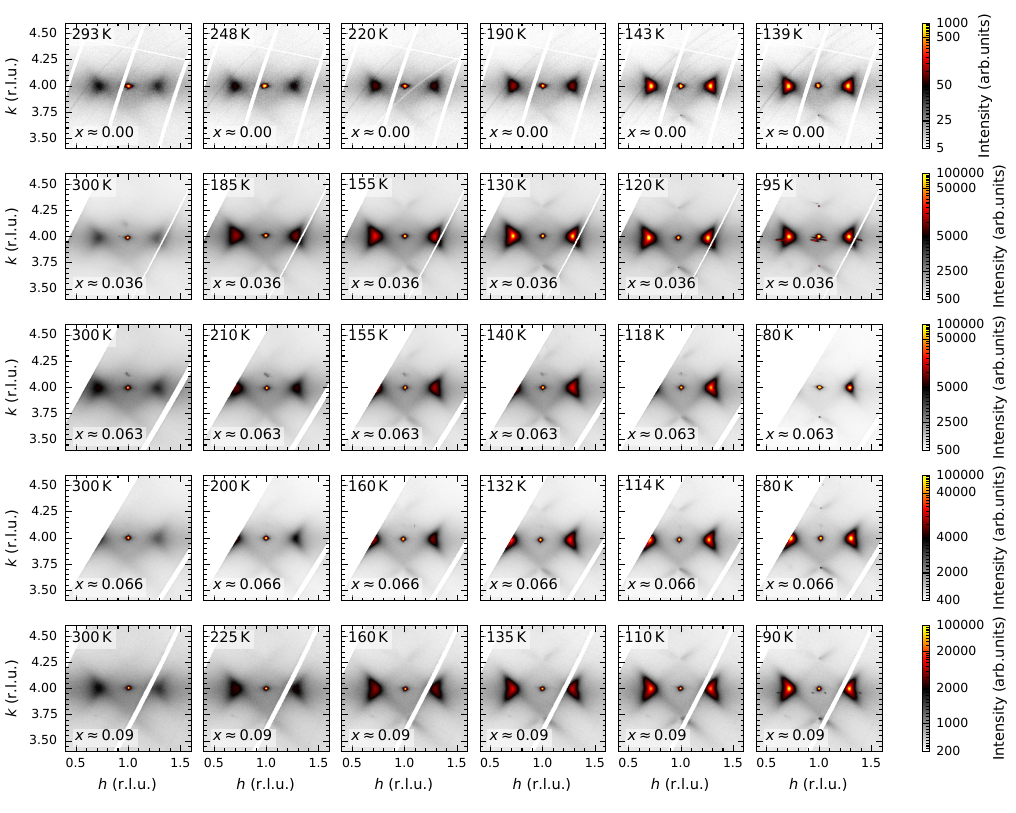}
\caption{\label{Fig:DS:HKMaps} \((h\ k\ 1)\) reciprocal space maps around the \((1\ 4\ 1)\) Bragg peak for \(x \approx 0; 0.036; 0.063; 0.066\) and \(0.09\) for different temperatures above the triclinic transition. The reciprocal space maps for \(x=0\) were reconstructed with data from \citet{Souliou2022}.}
\end{figure*}

In \cref{Fig:DS:HKMaps} \((h\ k\ 1)\) reciprocal space maps around the \((1\ 4\ 1)\) Bragg peak are shown for five different substitution levels. Due to the experimental limitations of the Displex cryostat, the \((1\ 4\ 1)\) Bragg was not accessible for the \(x\approx0.12\) sample and is not shown. For comparison  data for the \(x=0\) sample from \citet{Souliou2022} is also shown. Due to the differing measurement conditions and parameters, the absolute value of the intensity is different for those measurements. As discussed in the main text, the behavior is similar for all substitution levels. The diffuse scattering sharpens, gains in intensity and a sharp central peak is formed upon cooling. With increasing substitution level, only the transition temperature is lowered. Note that, for all substitutions, the peaks at the longitudinal direction (at low temperatures) and the diffuse scattering with the shape of a Rhombus are also present.

\begin{figure*}
\includegraphics{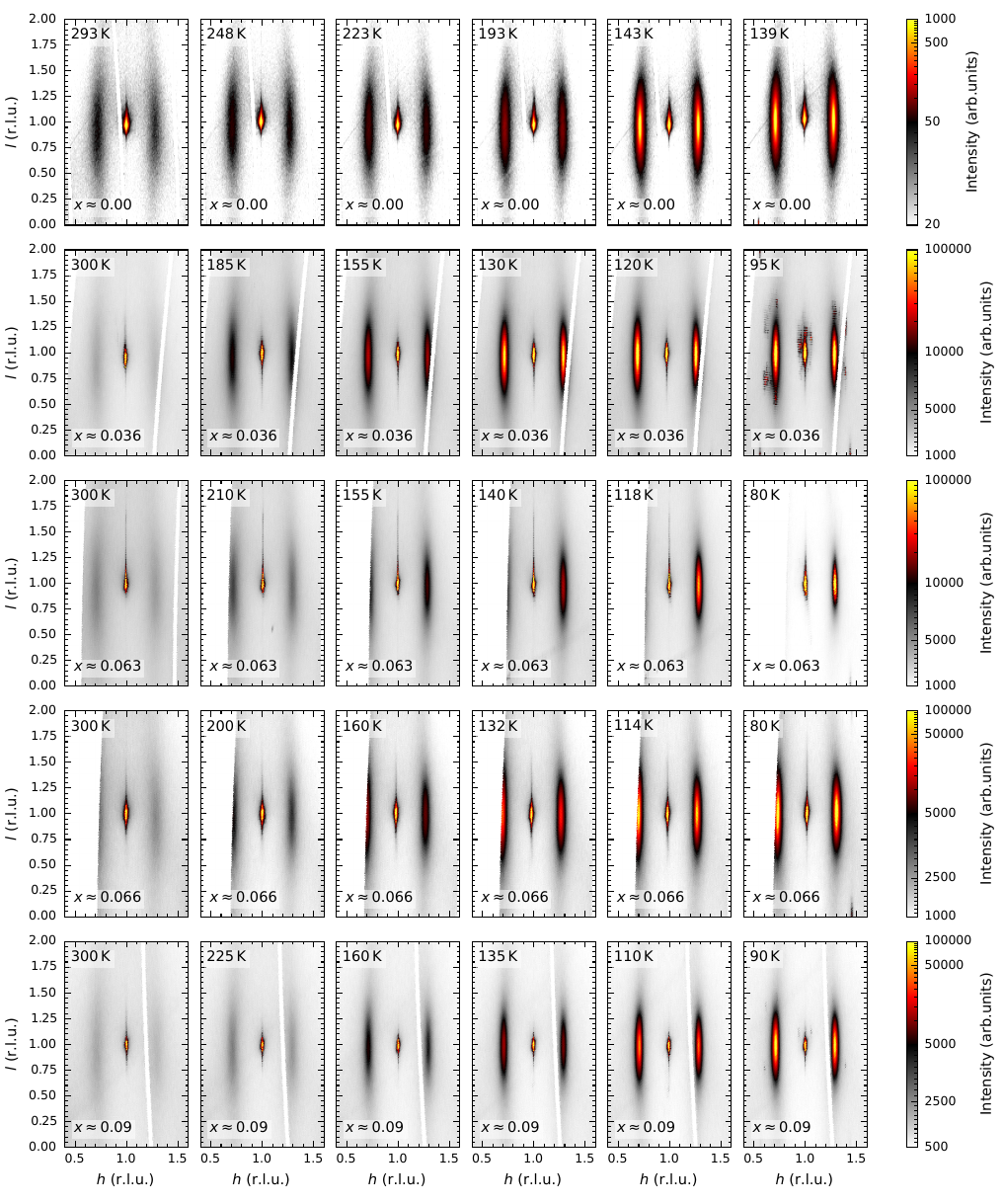}
\caption{\label{Fig:DS:HLMaps} \((h\ 4\ l)\) reciprocal space maps around the \((1\ 4\ 1)\) Bragg peak for \(x \approx 0; 0.036; 0.063; 0.066\) and \(0.09\)  for different temperatures above the triclinic transition. The reciprocal space maps for \(x=0\) were reconstructed with the raw data from \citet{Souliou2022}.}
\end{figure*}

Similar behavior is also observed in the \((h\ 4\ l)\) reciprocal space maps in \cref{Fig:DS:HLMaps}. The reconstructions show small artifacts from the reconstruction in the \(l\) direction due to the finite angular step size used for the data acquisition. However, the sharpening in the \(l\)-direction is more pronounced, as the correlation length is much smaller at room temperature. No special doping dependence or further features are visible in those maps.

\FloatBarrier
\subsection{Calculated three-dimensional diffuse scattering}

\begin{figure*}
\includegraphics{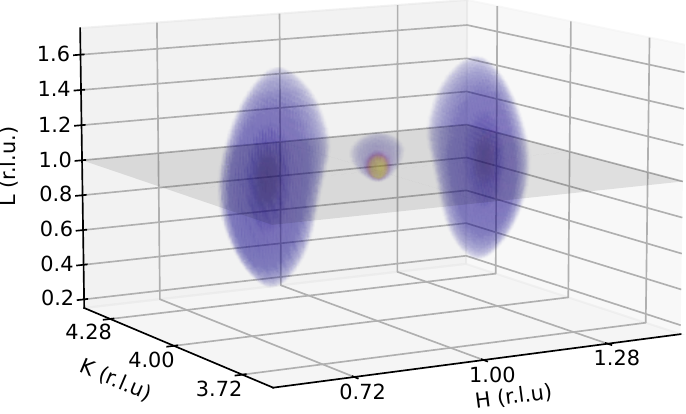}
\caption{\label{Fig:DS:Calc3D} Calculated three-dimensional diffuse scattering around the \((1\ 4\ 1)\) Bragg peak. For the temperature factors of the diffuse scattering, a temperature of \qty{102}{\kelvin} was used. }
\end{figure*}

\FloatBarrier
\subsection{I-CDW line cuts and fitting parameters}

\begin{figure*}
	\includegraphics{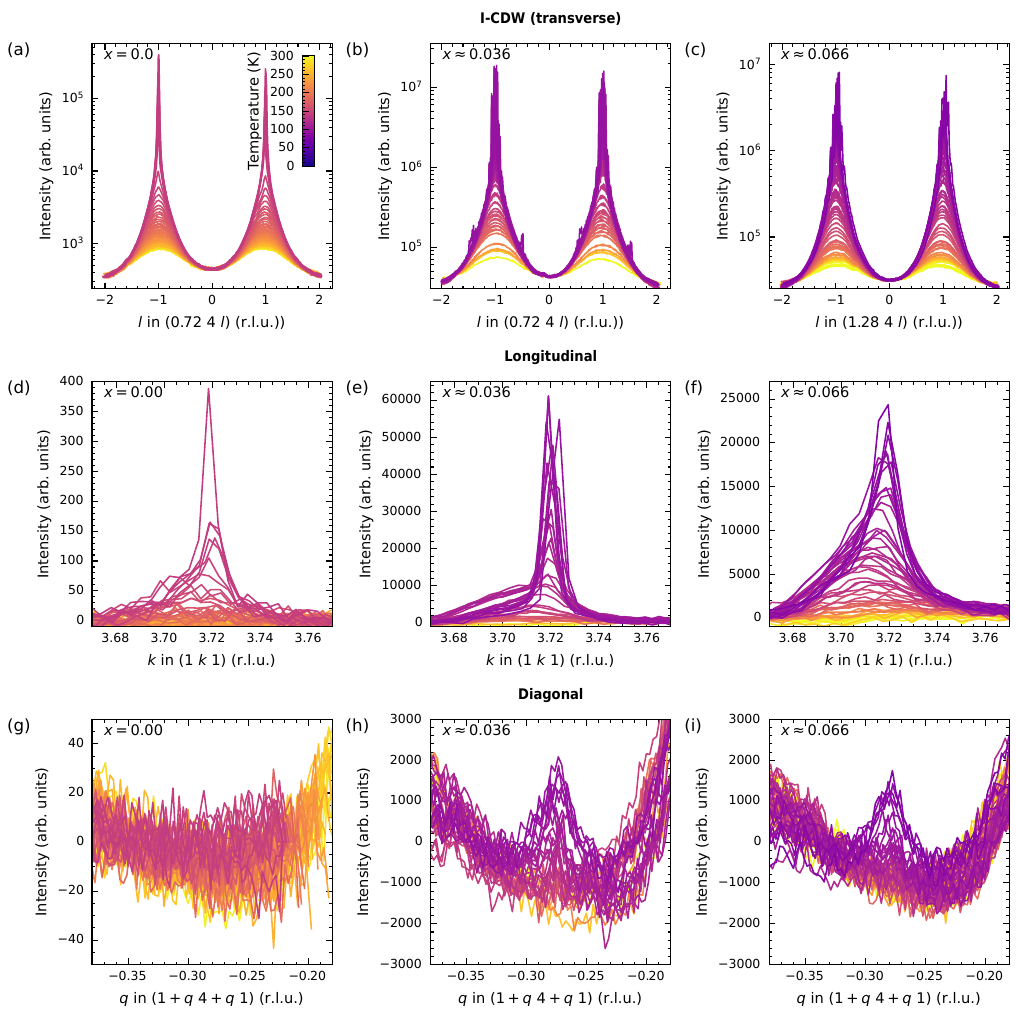}
	\caption{\label{Fig:DS:Cuts} (a)-(c) Linecuts in the \(l\) direction through the I-CDW peak at \((0.72\ 4\ 1)\) for different temperatures in the (a) \(x=0\), (b) \(x\approx0.036\) and (c) \(x\approx0.066\) samples. (d)-(f) Linecuts in the \(k\) direction through the longitudinal peak at \((1\ 3.72\ 1)\) for different temperatures in the (d) \(x=0\), (e) \(x\approx0.036\) and (f) \(x\approx0.066\) samples. (g)-(i) Linecuts in the \([1\ 1\ 0]\) direction through the longitudinal peak at \((0.72\ 3.72\ 1)\) for different temperatures in the (g) \(x=0\), (h) \(x\approx0.036\) and (i) \(x\approx0.066\) samples. Negative intensities result from imperfect background subtraction as described in the text. The linecuts for \(x=0\) were performed with data from \citet{Souliou2022}.}
\end{figure*}

In \cref{Fig:DS:Cuts} different line cuts through the I-CDW (transverse direction) and the peaks in the longitudinal direction and in the diagonal direction are shown. The linecuts are based on reciprocal space reconstructions as presented in \cref{Fig:DS:HLMaps,Fig:DS:HKMaps}. The cuts through the I-CDW are performed in the \(l\)-direction, while the other cuts are performed in-plane. The cuts through the I-CDW confirm the increase in intensity, sharpening and formation of a Bragg-like peak at low temperatures as discussed in the main text.

More importantly, in  \cref{Fig:DS:Cuts}d-f and g-i respectively, similar cuts through the longitudinal and diagonal peaks, which are much harder to observe in the reciprocal space maps, are shown. For visualization, a linear background is subtracted. While this is a good approximation for the longitudinal peaks, this is not the case for the diagonal peaks. Due to the lower intensity and proximity of the TDS signal along the rhombus, as discussed previously, the background is much stronger and challenging to subtract. The proper approach would be to fit the diffuse scattering from the rhombus and subtract the tails which extend to the position of the diagonal peaks. However, due to the very high difference in intensity, this is not reliable and can easily change the intensity of the peak. Furthermore, other background fits, such as polynomials, also severely alter the results. In the end, the most reliable method found was to subtract a linear background, even though the background is clearly non-linear.

Returning to the linecuts along the longitudinal direction, clear formation of a sharp peak is visible at low temperatures for all substitution levels. First, however, a diffuse and broad peak at slightly higher wavevector \(q\approx 0.3\) forms, which could be related to the rhombus and “tails” seen in the maps. Then, around the ordering temperature, a sharp peak emerges out of the diffuse background at the same wavevector as the I-CDW, but in the longitudinal, not transverse, direction. At a similar temperature, an even weaker peak at the diagonal direction \((0.28\ 0.28\ 0)\) emerges in the other linecuts. On one hand, the temperature dependence of the transverse (I-CDW), longitudinal and diagonal peaks is very similar, but, on the other hand, their absolute intensities are very different and the diagonal peak can only be observed in the phosphorous substituted samples. Nonetheless, except for these differences, all of the substitution levels exhibit a very similar behavior. Similarly to the I-CDW peak, only the temperatures and absolute intensities change with phosphorus substitution.

\begin{figure}
    \centering
        \includegraphics{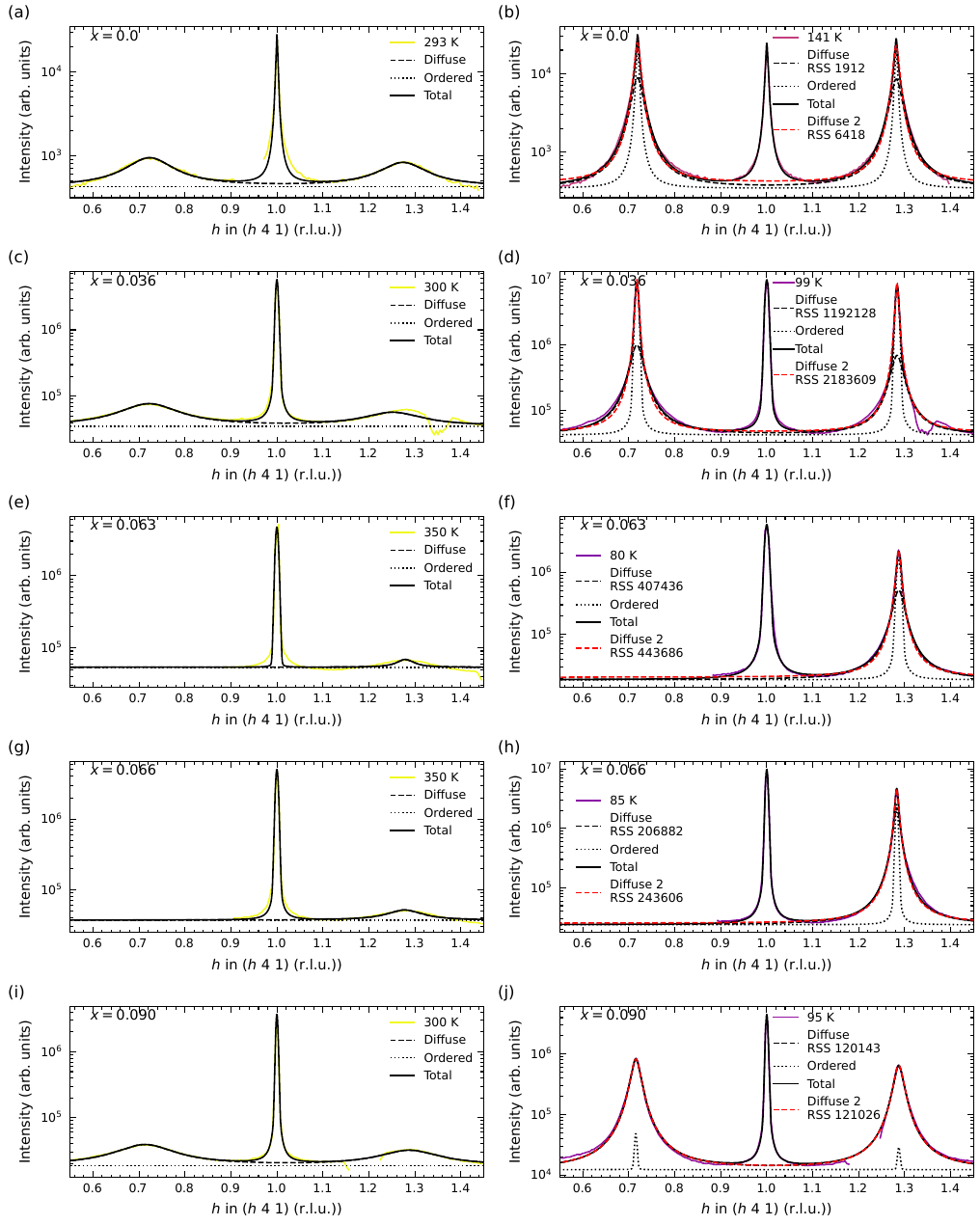}
        \caption{\label{Fig:TDSLineFits}  Fits of the Bragg peak and the I-CDW reflections in the \(h\)-direction are shown for different substitution contents at high and low temperatures. At low temperatures, the results of two fits are shown: one including the Pseudo-Voigt function (black) for the ordered peak at the I-CDW position, and one with only the Lorentzian function (red) for the diffuse scattering. For reference, the residual sum of squares (RSS) is given. The date for \(x=0.0\) are taken from \cite{Souliou2022}.}
    \centering
\end{figure}

\begin{figure}
    \centering
        \includegraphics{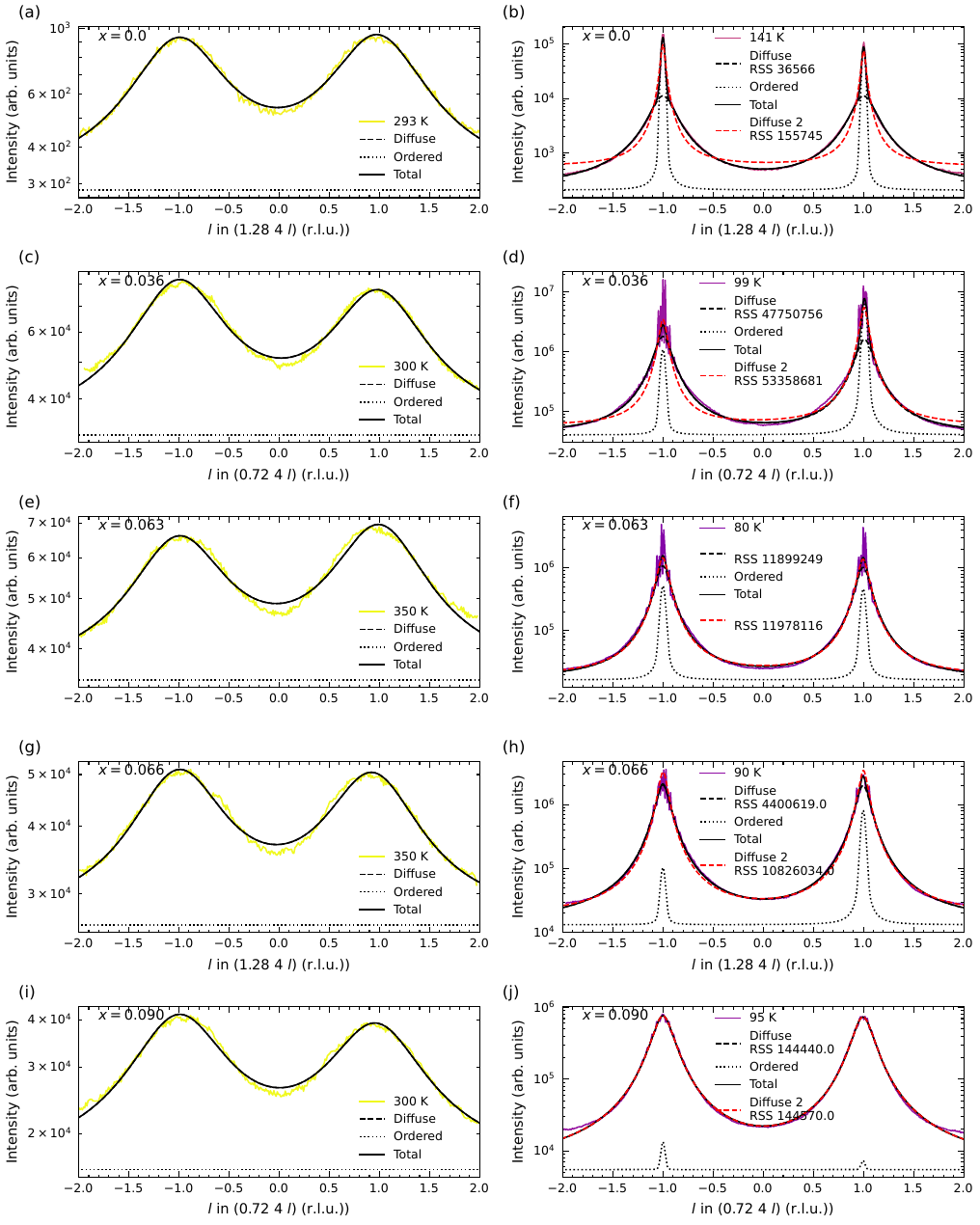}
        \caption{\label{Fig:TDSLLineFits}  Fits of the Bragg peak and the I-CDW reflections in the \(l\)-direction are shown for different substitution contents at high and low temperatures. At low temperatures, the results of two fits are shown: one including the Pseudo-Voigt function (black) for the ordered peak at the I-CDW position, and one with only the Lorentzian function (red) for the diffuse scattering. For reference, the residual sum of squares (RSS) is given. The date for \(x=0.0\) are taken from \cite{Souliou2022}.}
    \centering
\end{figure}

The I-CDW peaks were fitted to extract the intensity and the correlation length, as presented in the main text.  As mentioned in the main text, at low temperatures, a single Lorentzian line is not sufficient to fit the superstructure peak. Instead, a Lorentzian for the diffuse scattering must be combined with a Bragg-like pseudo-Voigt line with a shared center. \Cref{Fig:TDSLineFits,Fig:TDSLLineFits} show exemplary fits at high and low temperatures for different substitution contents. For reference, fits that do not include the pseudo-Voigt at low temperatures are also shown.

\begin{figure}
    \centering
        \includegraphics{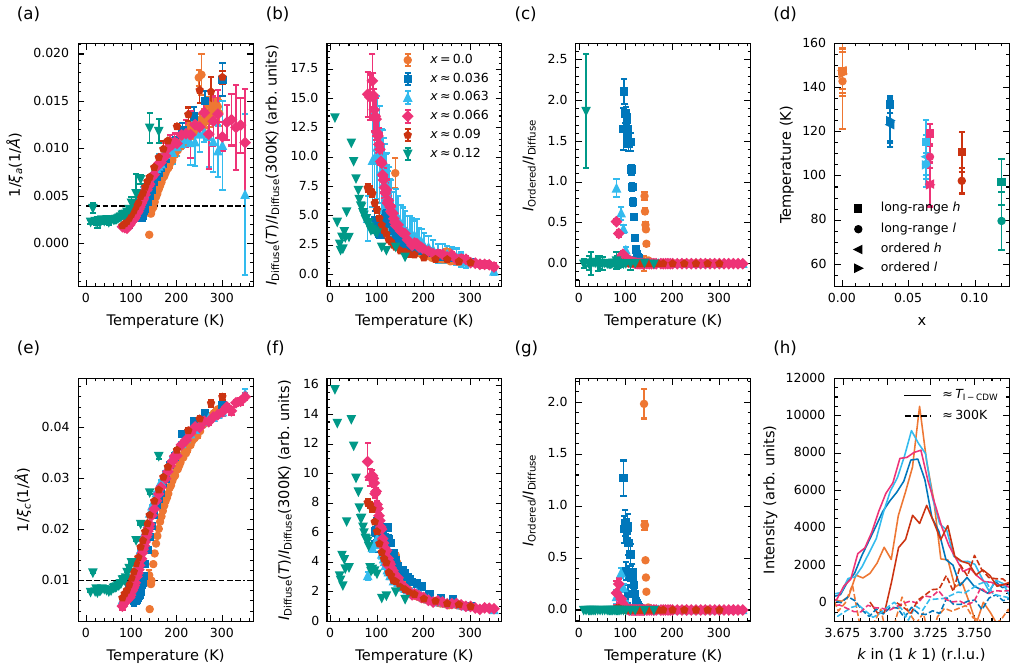}
        \caption{\label{Fig:TDSFits} (a), (e) Temperature dependence of the extracted inverse correlation lengths in the \(h\) and \(l\)-direction for different substitution levels, respectively. The dashed orange line indicates the critical correlation length extracted from the non-substituted sample. (b), (f) Temperature dependence of the intensity of the diffuse I-CDW peak in the (b) \(h\) and (f) \(l\)-direction. The data of the \(x\approx0.12\) sample is scaled due to the missing measurement at \qty{300}{\kelvin} to fit the data of the other substitution contents. (c), (g) Intensity ratio of the ordered and diffuse I-CDW peaks for different substitution levels extracted from linescans in the \(h\) and \(l\)-direction, respectively. (d) Extracted I-CDW transition temperature  from the critical correlation length (long-range) and the onset of the ordered intensity in the \(h\) and \(l\)-direction. (h) Linecuts through the longitudinal direction close to the I-CDW transition temperature as determined by the critical correlation length \(T_{\mathrm{I-CDW}}\) and at \qty{300}{\kelvin}}
    \centering
\end{figure}

All relevant fit parameters that are extracted from those fits are shown in \cref{Fig:TDSFits}. On cooling, the intensity of the diffuse scattering \(I_{\mathrm{Diffuse}}\) (the Lorentzian) increases significantly, by more than fifteenfold for certain samples. Initially there is a gradual increase in intensity before a rapid increase at the point where the correlation length also increases. Similarly to the correlation length, with increasing phosphorous substitution the change in intensity is shifted to lower temperatures and broadened.

The intensity of the Bragg like peak \(I_{\mathrm{Ordered}}\) behaves differently. For most temperatures the intensity is zero before it suddenly and very sharply gains intensity at the transition, eventually even more than the diffuse part. The behavior is very similar for the in-plane and out-of-plane directions and for all substitution levels. This onset of the ordered intensities can be used to define a criterion for the formation of the long-range ordered I-CDW, as discussed in the main text. To define the exact transition temperature of the static order, we use the crossing point of a linear fit of the ratio \(I_\mathrm{Ordered}/I_\mathrm{Diffuse}\) with 0. To ensure a reasonable fit, only points significantly different from 0 are selected, and the fit uses a weighting of the points by their error bars. The error state for the transition temperature is based on the statistical fitting error of this fit.

For at least the two lowest substitutions with the sharpest transition, the onset of the ordered intensity appear at a similar value of the correlation length. As discussed in the main text, this can be used to define an alternative criterion of the transition to a longe-range I-CDW fluctuation phase. To precisely define this temperature, the crossing point of a linear fit of the inverse correlation length and the inverse critical correlation length is used. This fit is restricted to points close to the critical correlation length and is weighted by their error bars. The stated error for this temperature is based on the statistical fitting error of the fit. The resulting temperatures are shown in \cref{Fig:TDSFits}(d).

As discussed previously, the longitudinal peaks also increase in intensity around the transition. Therefore, to cross-check the obtained transition temperatures, the linecuts at the first temperature below the extracted transition temperature are shown in \cref{Fig:TDSFits}h. The cuts from \qty{300}{\kelvin} are presented as a background. For all substitution levels, clear peaks have already formed. Together with the rather conservative choice of the critical correlation length, these temperatures should be viewed more as a lower bound of the transition temperature.

\begin{figure}
    \centering
        \includegraphics{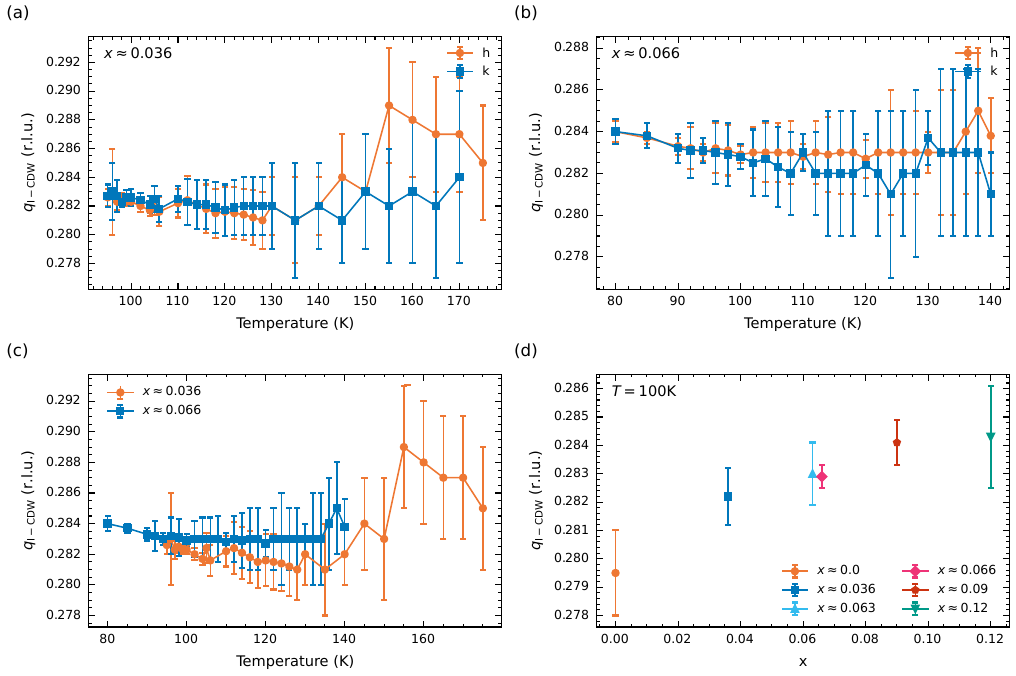}
        \caption{\label{Fig:TDSWV} (a)-(b) Temperature dependence of the I-CDW wavevector in the \(h\) direction for \(x\approx 0.036;0.066\). (c) Comparison of the temperature dependence of the I-CDW wave vector in the \(h\)-direction for the \(x\approx 0.036;0.066\) samples. (d) Phosphorus substitution dependence of the I-CDW wavevector at \qty{100}{\kelvin}. The non-substituted sample wavevector is taken at \qty{139}{\kelvin} and based on data from \citet{Souliou2022}.}
    \centering
\end{figure}

In addition to the correlation length and intensity, the wavevector can also be extracted by investigating the peak position of several I-CDW peaks in a large volume of reciprocal space. An example temperature dependence is shown in  \cref{Fig:TDSWV} (a)-(b) for two substitution levels. While the temperature dependence remains tiny and is comparable for different substitution levels, the wavevector increases with higher phosphorous content, as indicated in \cref{Fig:TDSWV}(d).

\begin{figure}
    \centering
        \includegraphics{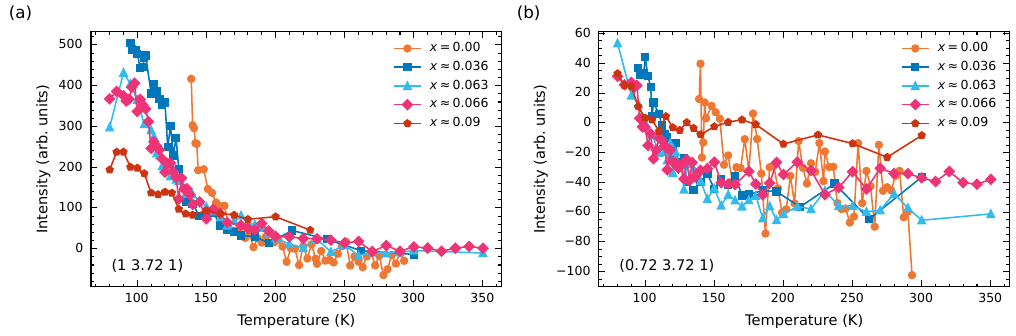}
        \caption{\label{Fig:TDSIntPeaks} (a) Integrated intensity of the linecuts in the longitudinal direction versus temperature for the different substituted samples. (b) Integrated intensity of the linecuts in the diagonal direction versus temperature for the different substituted samples. Negative intensities result from imperfect background subtraction as described in the text.}
    \centering
\end{figure}

As the intensity for the peaks in the longitudinal and especially the diagonal directions is significantly weaker, a fit of the data would not be very reliable. In the raw data, the peaks seem as sharp as a Bragg peak, and therefore the fitted width would not be insightful. However, the data can still be integrated to determine the integrated intensity of the peaks. The results from this calculation are shown for the longitudinal and diagonal directions in \cref{Fig:TDSIntPeaks}. Both peaks follow a similar trend of showing no or only a slight increase in intensity down to the ordering temperature. Below this, a clear increase in intensity is observed. For higher substitution levels, the increase is more shallow.

\FloatBarrier
\section{I-CDW IXS}

\subsection{Comparison of none- and phosphorus-substituted IXS scans at the I-CDW position}
\begin{figure*}
\includegraphics{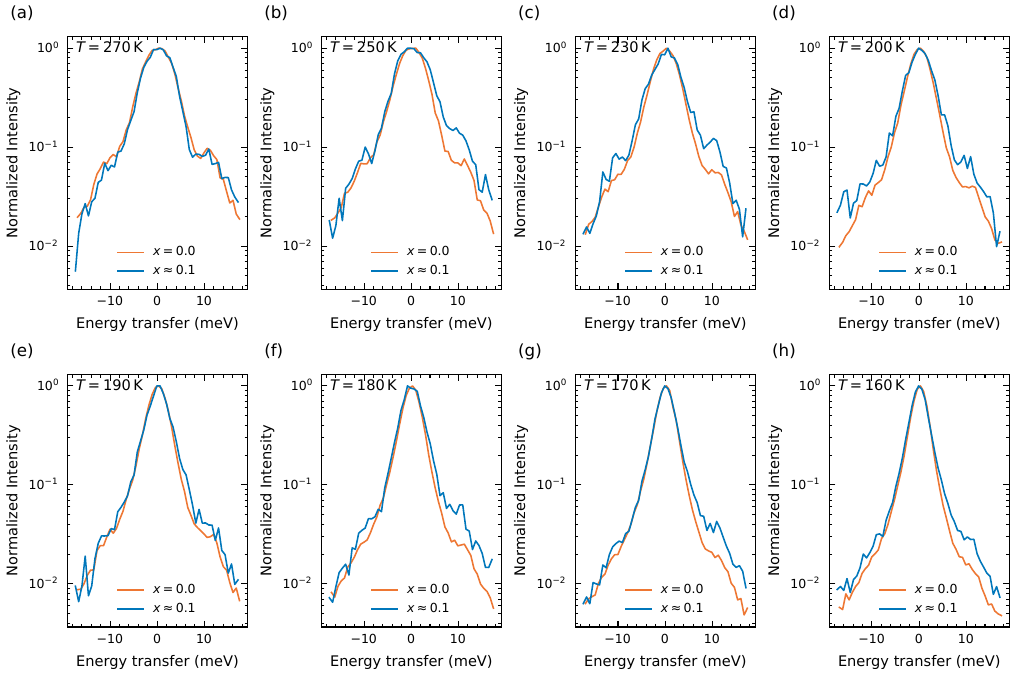}
\caption{\label{Fig:IXS:Comp} One-to-one comparison of the IXS spectra at the I-CDW position of the \(x=0\) and \(x\approx0.1\) sample for (a) \qty{270}{\kelvin}, (b) \qty{250}{\kelvin}, (c) \qty{230}{\kelvin}, (d) \qty{200}{\kelvin}, (e) \qty{190}{\kelvin}, (f) \qty{180}{\kelvin}, (g) \qty{170}{\kelvin} and (h) \qty{160}{\kelvin}. The data for \(x=0\) was reproduced from \citet{Souliou2022}.}
\end{figure*}

As discussed in the main text, at \qty{160}{\kelvin} the spectrum of the \(x\approx0.1\) sample is broader than for \(x=0\) at the I-CDW position, indicating a higher phonon energy. A one-to-one comparison of the spectra at the I-CDW position for all available temperatures is shown in \cref{Fig:IXS:Comp}. For the comparison, the spectra are normalized to compensate for different absolute intensities. In agreement with the discussion in the main text and the results from the TDS measurements at all temperatures, the spectrum for the substituted sample is broader, indicating the lower transition temperature of the phosphorus substituted sample.

\FloatBarrier

\subsection{Not normalized IXS spectra and fit of the dispersion at \qty{170}{\kelvin}}
\begin{figure*}
\includegraphics{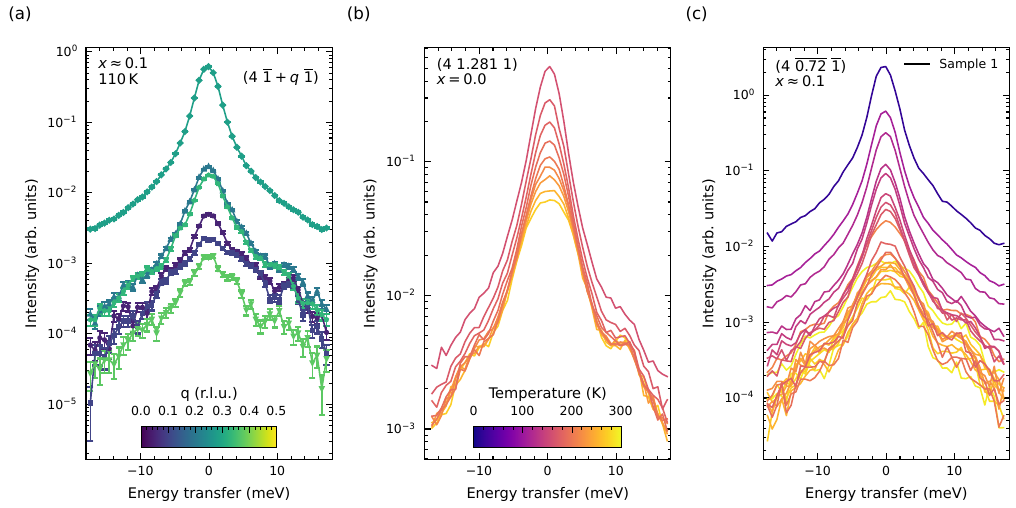}
\caption{\label{Fig:IXS:NoNorm} (a) IXS spectra along the [0 1 0]-direction around the
I-CDW wave vector at \qty{110}{\kelvin}. (b)-(c) Temperature dependence of the IXS raw spectra at the I-CDW position for (b) the \(x=0.0\) and (c) the \(\approx 0.1\) samples.}
\end{figure*}

In the main text the most IXS spectra are normalized to compensate for the strong intensity change due to the phonon softening. In \cref{Fig:IXS:NoNorm} we present the normalized IXS spectra from the main text without normalization.

The IXS spectra consist of one resolution-limited elastic line and two damped harmonic oscillator functions convoluted with the resolution function for the two phonons. The damped harmonic oscillator functions are further weighted by the temperature dependent Bose factor. To restrict the fits of the dispersion at low temperatures the phonon structure factor is kept constant throughout the dispersion, as it structure factor is nearly constant over the investigated momentum range \cite{Souliou2022}. For example the fit of the dispersion at \qty{170}{\kelvin} is shown in \cref{Fig:IXS:FitDisp170K}.

\begin{figure*}
\includegraphics{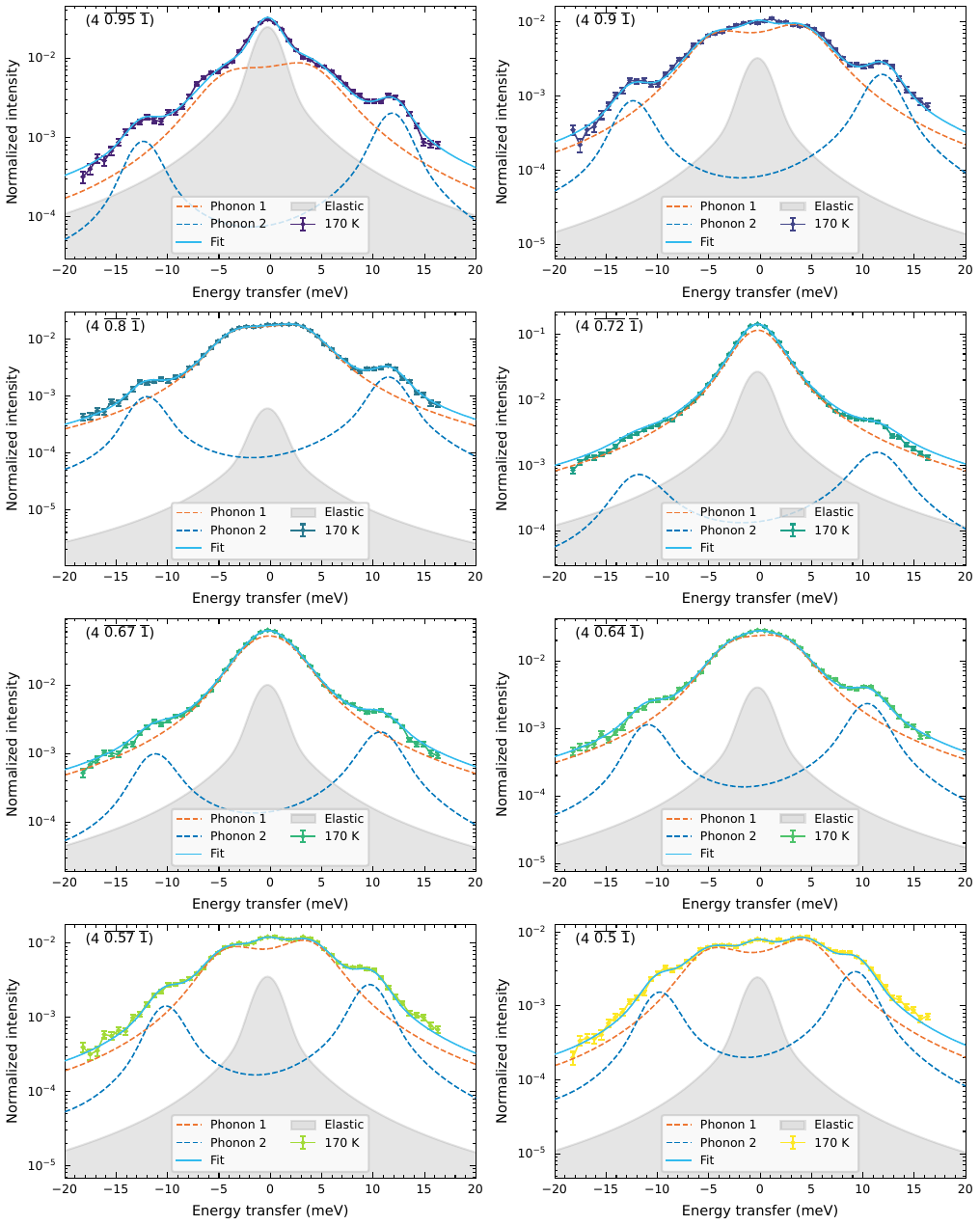}
\caption{\label{Fig:IXS:Fig:IXS:FitDisp170K} IXS spectra and fit of the elsatic line and phonons for the dispersion at \qty{170}{\kelvin}. The individual components of the fit are shown.}
\end{figure*}

\FloatBarrier
\subsection{Phonon dispersion around the superconducting transition temperature}

\begin{figure*}
\includegraphics{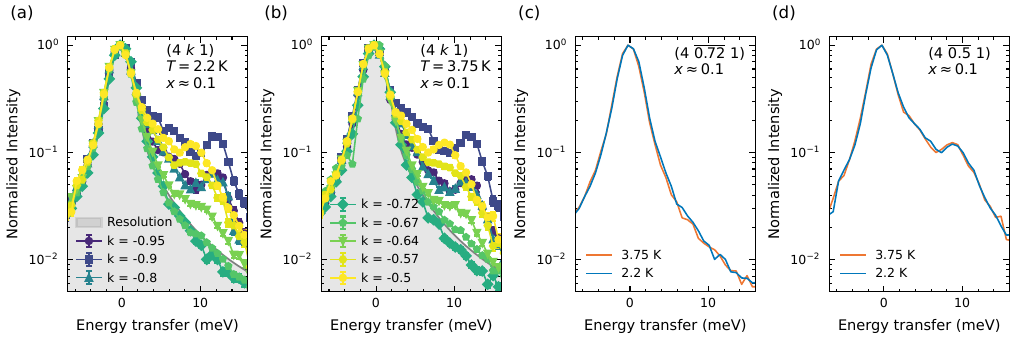}
\caption{\label{Fig:IXS:Tc} (a)-(b) normalized IXS spectra along the I-CDW direction around the \((4\ 1\ 1)\)-Bragg peak (a) (nominally) below the superconducting transition and (b) above the superconducting transition. (c)-(d) one-to-one comparison of the normalized IXS spectra at (c) the I-CDW position \((4\ \overline{0.72}\ 1)\) and (d) \((4\ \overline{0.5}\ 1)\). The data at \((4\ \overline{0.72}\ 1)\) was measured with an absorber in the incoming beam. The data were measured on a \(x\approx 0.1\) sample.}
\end{figure*}

The Joule-Thompson displex cryostat used for the measurements of the \(x\approx0.1\) sample had a base temperature of \qty{2.2}{\kelvin} at the time of the experiment. As the superconducting \(T_c\) is \qty{3.2}{\kelvin} for the \(x\approx0.1\) substitution level, this allows, nominally, for measurements in the superconducting phase in order to investigate the interplay between superconductivity and the I-CDW. Nonetheless, one should keep in mind that the temperatures are nominal temperatures. A simultaneous resistivity measurement or a thermometer on the sample/very close to the sample, in order to unambiguously confirm the superconducting transition, was not possible due to experimental limitations. Furthermore, ideally, the measurements would have been performed at temperatures \(T<1/3T_c\).

The spectra along the I-CDW-direction at \(\qty{2.2}{\kelvin}<T_c\) and \(\qty{3.75}{\kelvin}>T_c\) are presented in Figs. \ref{Fig:IXS:Tc}(a)-(b). The spectra for both temperatures look nearly identical. This is also confirmed by the one-to-one comparison at the I-CDW position and the last point measured in Figs. \ref{Fig:IXS:Tc}(c)-(d). Therefore, within the limitations described above, no competition between I-CDW and superconductivity can be identified.

\FloatBarrier
\section{I-CDW IXS along longitudinal direction} \label{Sec:IXSLong}
\begin{figure*}
	\includegraphics{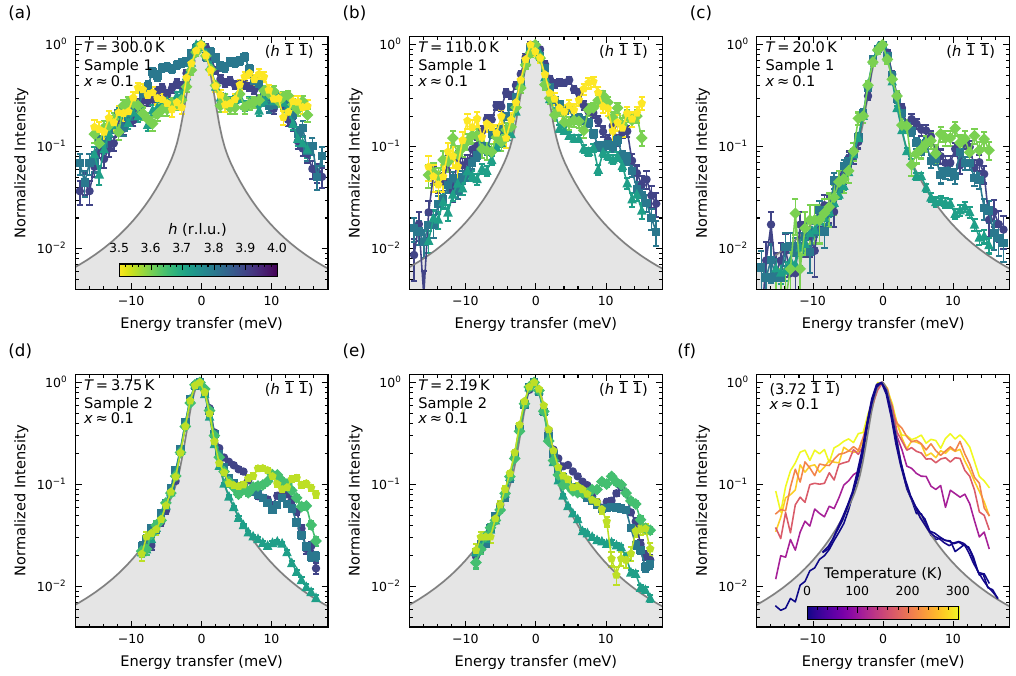}
	\caption{\label{Fig:IXS:ICDWOrthRaw} Normalized IXS spectra long the \([h\ 0\ 0]\)-direction around the \((4\ \overline{1}\ \overline{1})\)-Bragg peak at (a) \qty{300}{\kelvin}, (b) \qty{110}{\kelvin}, (c) \qty{20}{\kelvin}, (d) \qty{3.75}{\kelvin} and (e) \qty{2.2}{\kelvin}. (f) Normalized IXS spectra at \((3.72\ \overline{1}\ \overline{1})\) for different temperatures. The data were measured on two different samples \(x\approx 0.1\).}
\end{figure*}

The TDS measurements indicate the formation of a peak in the longitudinal direction at low temperatures. To investigate the origin of this peak, IXS scans in the longitudinal direction and a temperature dependence at the position of the peak are shown in \cref{Fig:IXS:ICDWOrthRaw}. Similarly to the results of \citet{Souliou2022} in the pristine compound, no soft phonon could be identified. Instead, at low temperatures, a strong increase in the elastic line with no or only little change to the phonons can be seen. This confirms that the peak observed in TDS in the longitudinal direction is static and most likely a signature of the static I-CDW order.

\FloatBarrier
\section{High-pressure instability position --- I-CDW2}

\begin{figure*}
\includegraphics{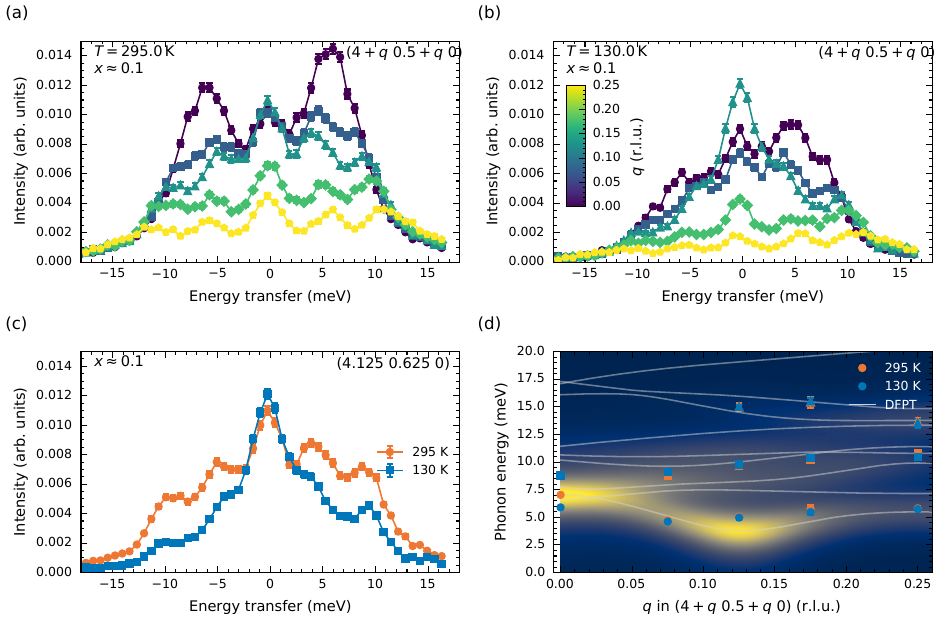}
\caption{\label{Fig:IXS:HPDisp} (a)-(b) IXS spectra at \((4+q\ \frac{1}{2}+q\  0)\) at (a) \qty{300}{\kelvin} and (b) \qty{130}{\kelvin}. (c) direct comparison of the spectra at \((4.125\ 0.625\ 0)\) at \qty{300}{\kelvin} and \qty{130}{\kelvin}. (d) fitted phonon energies at \qty{300}{\kelvin} (orange) and \qty{130}{\kelvin} (blue). The white lines indicate the phonon dispersion calculated by DFPT and the background the phonon from factor smeared. For visualization, the IXS intensity is broadened by a Gaussian with a FWHM of \qty{0.7}{\milli\eV}.}
\end{figure*}

As shown by \citet{Lacmann2023}, first-principles calculations show a phonon anomaly close to the position of a second I-CDW (denoted as I-CDW2) already at low pressure. In fact, it turns out that DFPT calculations using the experimentally-determined tetragonal crystal structure at a substitution of \(x=0.1\) and ambient pressure also exhibit signatures of the anomaly. Note, however, that the DFPT calculations were not performed with an actual exchange of phosphorus atoms for arsenic atoms, as this is generally difficult to do in DFPT. Together with the diffuse scattering signal around the same positions, this indicates that there could also be signatures of this instability at ambient pressure in the \(x\approx0.1\) sample. 

The measured IXS spectra and the fitted dispersion are shown in \cref{Fig:IXS:HPDisp}. The spectra along the \((4+q\ 1/2+q\ 0)\)-line do not show an anomaly, neither at \qty{295}{\kelvin} nor at \qty{130}{\kelvin} -- below the I-CDW2 ordering temperature in the non-substituted samples \cite{Lacmann2023}. The spectra do not change significantly between \qty{295}{\kelvin} and \qty{130}{\kelvin}. This is also true for the point closest to the I-CDW2 position in \cref{Fig:IXS:HPDisp}(c), where the difference is mainly because of the different phonon occupation due to the Bose-Einstein distribution. It should be noted that the position of the I-CDW2 used here is an approximation based on the observation of the I-CDW2 under hydrostatic pressure for other substitution levels \cite{Lacmann2023} and the results of the DFPT calculations. The real position of the softening for this temperature and substitution level is not known and is difficult to determine, as the measurements under hydrostatic pressure show a strong dependence of this position on both pressure and temperature and the DFPT calculations suggest a slightly different wave vector \cite{Lacmann2023}. Especially if the position of the phonon softening is sharp in reciprocal space, it is possible that the anomaly was just missed. However, in the complete dispersion shown in \cref{Fig:IXS:HPDisp}(d) there is no clear anomaly and little change between the two temperatures. This suggests that the DFPT calculations overestimate the I-CDW2 instability, similar to as observed for the TDS measurements in the main text. Nevertheless, the measured dispersion overall shows a good agreement with the DFPT calculations. The small difference between the measured and calculated phonon dispersion could come from close-lying phonon branches and the limited energy resolution.

\begin{figure*}
	\includegraphics{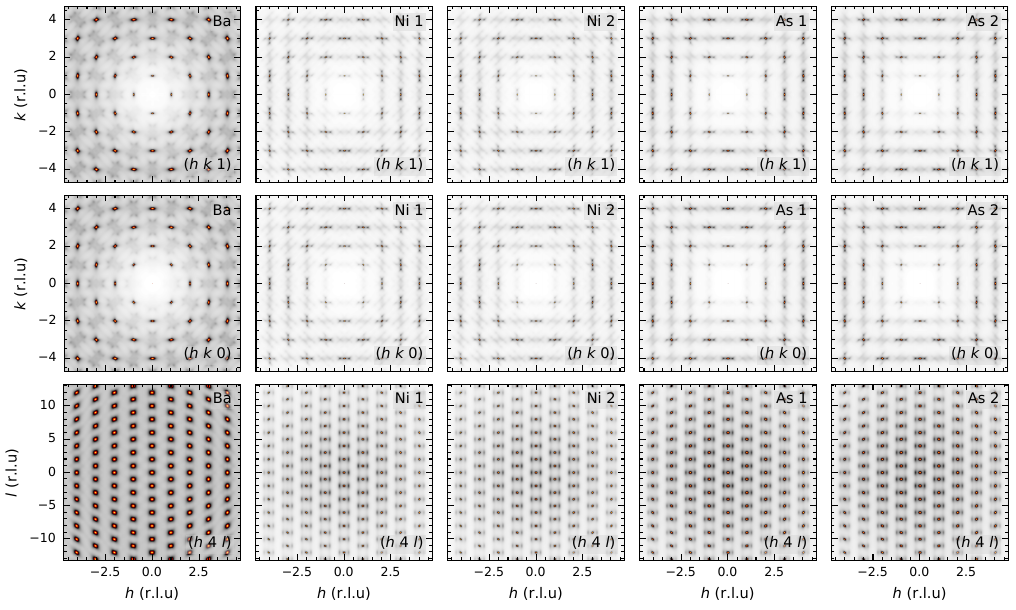}
	\caption{\label{Fig:DS:Calc} Calculated \((h\ k\ 1)\), \((h\ k\ 0)\) and \((h\ 4\ l)\) reciprocal spaces maps of the TDS with the temperature factor for \qty{102}{\kelvin}. The maps show the contribution from the barium, the two nickel and two arsenic atoms individually.}
\end{figure*}

The DFPT calculations also show a major difference between the I-CDW and I-CDW2. As shown in the maps in \cref{Fig:DS:Calc}, the diffuse scattering at the I-CDW2 position involves all the possible atom types. This is different to the I-CDW, for which only the Ni and As atoms are involved. Furthermore, scattering from no single type of atom shows the complete structure observed in the TDS measurements and only the combination of all atoms gives the right pattern. As the patterns from single atoms show some kind of asymmetry, this might also explain the distortion of the I-CDW2 observed at high pressures, as the different kinds of atoms may react differently to the applied pressure.

\FloatBarrier
\bibliography{IXSDS2024}